\documentstyle[twocolumn,prb,aps]{revtex}
\input psfig
\begin{document}
\draft
\twocolumn[\hsize\textwidth\columnwidth\hsize\csname@twocolumnfalse\endcsname
\title{Non--Equilibrium Dynamics of the Anderson Impurity Model}
\author{Matthias~H.~Hettler,$^{1*}$ Johann Kroha,$^{2\dag}$ and 
Selman Hershfield$^{1\ddag}$}
\address{$^1$ Department of Physics and National High Magnetic Field 
Laboratory,
University of Florida, Gainesville, FL 32611}
\address{$^2$ Institut f\"ur Theorie der Kondensierten Materie,
Universit\"at Karlsruhe, 76128 Karlsruhe, Germany}
\date{Received \today}
\maketitle
\begin{abstract}
The $M$--channel Anderson impurity model ($M=1,2$) is studied in the 
Kondo limit with a finite voltage bias applied 
to the conduction electron reservoirs. 
Using the Non--Crossing Approximation (NCA),
we calculate the local spectral functions, the differential conductance, 
and susceptibility at non--zero bias for symmetric as well as
asymmetric coupling of the impurity to the leads. 
We describe an effective procedure 
to solve the NCA integral equations which enables us to reach 
temperatures far below the Kondo scale. This allows us to study
the scaling regime where the conductance  depends on the bias 
only via a scaling function.
Our results are applicable to both tunnel junctions 
and to point contacts. 
We present a general formula which allows one to go between the
two cases of tunnel junctions and point contacts.
Comparison is also made between the conformal field theory 
and the NCA conduction electron self-energies in the two channel case.
\end{abstract}
]
    
\narrowtext
\section{Introduction}
In recent years, the Kondo model and the Anderson impurity model in
its Kondo limit have been investigated extensively by use 
of numerical renormalization group (NRG) calculations,\cite{wilson,oliveira} 
the Bethe ansatz method,\cite{betheansatz,bethemulti}
conformal field theory (CFT),\cite{afflud} and auxiliary particle 
techniques. In this way a consistent
theoretical understanding of the Kondo effect in equilibrium has emerged. 
In particular, the ground state of the system
depends on the symmetry group of the conduction electron
system: If the number of channels, $M$, is less than the level degeneracy,
$N$, the screening of the local moment at 
energies below the Kondo scale, $T_K$, leads to a singlet Fermi 
liquid ground state with strongly renormalized Fermi 
liquid parameters. If, in contrast, $M\geq N$, the ground state 
is M--fold degenerate, leading to a non-vanishing entropy at zero temperature $T$
and a characteristic energy dependence of the density of 
states, obeying fractional power law below the
Kondo scale. This is mirrored in an anomalous (non--Fermi liquid)
behavior of the thermodynamics as well as transport properties\cite{afflud}.

On the other hand, there has been much less work on the non--equilibrium Kondo
problem, where the electron distribution is not in local equilibrium 
about the Kondo impurity and linear response theory is no longer sufficient.
Possible new effects in this situation include the breaking of time reversal 
symmetry and the appearance of a new energy scale like the charge transfer 
rate through a tunneling or point contact.
The phenomena of tunneling through magnetic impurities has been
explored  since the 1960's when zero bias anomalies (ZBA) were observed in
metal-insulator-metal tunnel junctions.\cite{zbaexp,zbamag}
The origin of these
zero bias anomalies was understood in terms of perturbative theories,
\cite{app-and}
which captured the basic phenomena: a logarithmic temperature dependence
and a Zeeman splitting of the ZBA peak in a finite magnetic field.
Although the original theories were quite successful in fitting the
data,\cite{app-shen} they were not able to get to what we now know as the low
temperature strong coupling regime of the Kondo problem.
In view of theoretical advances since that time, it is 
worthwhile to re-examine the non--equilibrium Kondo effect,
particularly in the low temperature regime.

In addition, there have been a number of new and interesting
realizations of the Kondo effect in non--equilibrium.  
With recent advances in sample fabrication, it has become
possible to see a zero bias anomaly caused by a single Kondo 
impurity.\cite{rabu2}  Even more intriguing is the observation
of zero bias anomalies in point contacts that exhibit
logarithmic temperature dependence
at high temperatures and power law behavior at low temperatures
but no Zeeman splitting in a magnetic field \cite{rabu,rlvdb,Upadhyay}.  
Such zero bias anomalies may be described by the
two--channel Kondo model, where the
ZBA is caused by
electron assisted tunneling in two--level systems (TLS),
although other descriptions have been proposed as well.
\cite{win-meir-alt}

In the 1980's Zawadowski and Vladar \cite{zv1} showed that
if two-level-systems with sufficiently
small energy splittings existed in metals, then one could
observe a Kondo effect due to the electrons scattering off
these TLS.  In this case the TLS plays the role of a pseudo--spin. 
One state of the two-level-system
may be regarded as pseudo--spin-up and the other as pseudo--spin-down.
An electron scattering off  the TLS can cause the 
pseudo-spin state of the TLS to change. The electron state also changes, e.g. 
its parity, in the process. This electron--assisted pseudo--spin--flip
scattering plays the role of spin-flip scattering in the
standard case of the magnetic Kondo effect.
Detailed analysis, taking into account
the different partial waves for scattering off the TLS
shows that a Kondo effect is indeed generated by this electron--assisted 
tunneling. However, since the true electron
spin is conserved in scattering from the TLS,
there are two kinds or channels of electrons. Hence the
the system will display the two--channel Kondo effect\cite{nozbl}.
Level splitting and multi--electron scattering may disrupt the
two--channel non--Fermi liquid behavior.
The stability of the two--channel fixed point against these
perturbations is currently a subject of 
investigation\cite{zv1,zarand,moustakas}.\\

In recent years a number of techniques have been applied to the
non--equilibrium Kondo problem: variational calculations,\cite{ng}
perturbation theory,\cite{sdw}
equation of motion,\cite{mewinlee} perturbative functional  
integral methods,\cite{Schoeller} 
and exactly solvable points of the model.\cite{schill-hersh}
One of the most powerful techniques in this context is the
auxiliary boson technique\cite{barnes,keiter,cole}.
It has two major advantages:
(i) In its lowest order self--consistent approximation, the 
non--crossing approximation (NCA)\cite{kuram,muhart,bickers}, it 
yields an accurate quantitative description of the single--channel Anderson
model in equilibrium\cite{bickers,costi,coxzawa} 
down to low temperatures, although it does not capture correctly the
Fermi liquid regime.  The NCA even describes
the infrared dynamics of the two--channel model correctly as one approaches
zero temperature.
\cite{cox3} (ii) The NCA is based on a standard self--consistent
Feynman propagator expansion. Therefore, in contrast to exact solution
methods, it need not rely on special symmetry properties which are not
always realized in experiments. The formalism
also allows for systematic improvements 
of the approximation.\cite{kroha2,kroha,anders} 
Moreover, the NCA may be generalized for
non--equilibrium cases in a straight--forward manner. 
This has been achieved recently for
the single--channel Anderson model.\cite{mewin2,mewin3} However,
the low temperature strong coupling regime of the model was not reached. 

In this article we give a formulation of the NCA 
away from equilibrium which allows for a highly efficient numerical 
treatment, so that temperatures well inside the low energy  
scaling regime may be reached.
In order to enable other researchers in this field to more readily
apply this method to related problems we describe the numerical
implementation of the formalism in some detail. 
Subsequently, we 
study a number of non-equilibrium properties of the single-- and especially 
the two--channel Anderson model in the Kondo regime: 
(1)
In linear response we study the conductance of the
two channel model and the one
channel model for different spin degeneracies.  We use both
tunnel junction and point contact geometries and discuss
how to go continuously between the two.  These results are
compared to the bulk resistivity.
(2)
The nonlinear response is computed for the same Anderson models,
and the scaling of the differential conductance at low temperatures
and voltages is studied.
(3) 
The NCA self-energies are compared to those
obtained by conformal field theory. This sheds light on the 
question how far the
equilibrium CFT results for the scaling function are applicable to
non--equilibrium situations.
(4)
The effect of an asymmetry in the coupling of the impurity to
the two leads is studied and shown to be consistent with asymmetries
of ZBA's observed experimentally.
(5) 
Finally, we compute the effect of finite bias on the local (pseudo)spin 
susceptibility.
Temperature and voltage scaling is
verified below $T_K$, but large differences between the
temperature and the voltage dependence are found outside of the 
scaling regime.\\

The paper is organized as follows: In section II the one-- and
two--channel Anderson models and their applicability to tunnel junctions and
point contacts with defects are discussed. Section III 
contains the formulation of the problem within the NCA and discusses its
validity for the single-- and the two--channel case, respectively. 
An effective method for the solution of the NCA equations both for 
equilibrium and for static non--equilibrium is introduced. 
Section IV contains the results for the quantities
mentioned above, which are discussed in comparison with equilibrium 
CFT solutions and experiments, where applicable. 
All the results are summarized in Section V.
\section{The SU(N)$\times$SU(M) Anderson impurity model out of equilibrium}
\subsection{The model and physical realizations}
The single--channel ($M=1$) and multi--channel ($M>1$) Kondo effects
occur when a local $N$--fold degenerate degree of freedom,
$\sigma=1,\dots ,N$, is coupled via an exchange interaction to $M$
identical conduction electron bands, 
characterized by a continuous density of states
and a Fermi surface. 
For example, the ordinary Kondo effect occurs when a magnetic impurity
is coupled to conduction electrons via an exchange interaction.
The impurity with spin $S$ plays the role of the $N=2S+1$
degenerate degrees of freedom, and there is only one flavor 
or channel of conduction electrons so $M=1$.
There are other physical situations where there are $M$
bands of conduction electrons which are not
scattering into each other.  In this case one says that there
are $M$ channels or flavors of electrons.  
For the $M$ channel Kondo effect,
the channel or flavor degree of freedom,
$\tau = 1,\dots ,M$, is assumed to be conserved by the exchange coupling.

Because of the
non--canonical commutation relations of the spin algebra, this model is
not easily accessible by standard field theoretic methods. 
Rather than work directly with the Kondo model, it is frequently
more convenient to work with the corresponding Anderson model.
Within the Anderson model, each of the possible
spin or pseudo-spin states, $\sigma$, is represented
by a fermionic particle. By convention the operator
which creates a fermion in the local level $\sigma$ from a conduction
electron in channel $\tau$ is denoted by $d_{\sigma \tau}^{\dag}$.
Since each of the d--states created by $d_{\sigma \tau}^{\dag}$,
represents a different (pseudo)spin state, only one of the states
should be occupied at a time. In order to enforce this
constraint, we use the auxiliary boson technique\cite{barnes} and write
$d_{\sigma \tau}^{\dag}$ as
$d^{\dag}_{\sigma\tau}=f^{\dag}_{\sigma}b_{\bar\tau}$,
where $f_{\sigma}$ is a fermion operator and
$b_{\tau}$ is a boson operator describing the unoccupied local d--level.
The constraint is then written as the operator identity
$Q=\sum _{\sigma}
f^{\dag}_{\sigma}f_{\sigma}+\sum _{\tau}b^{\dag}_{\tau}b_{\tau}=1$. 
In terms of pseudo-fermion operators $f_{\sigma}$ and slave boson operators
$b_{\bar\tau}$, the SU(N)$\times$SU(M) Anderson model is 
\begin{eqnarray}
H&=&\sum _{\vec k,\sigma,\tau,\alpha}(\varepsilon _{\vec k}-eV_{\alpha}) 
c^{\alpha \ \dag}_{\vec k\sigma\tau}  c^{\alpha}_{\vec k\sigma\tau}+
\varepsilon _d \sum _{\sigma} f^{\dag}_{\sigma } f_{\sigma} \nonumber\\ 
&+& \sum _{\vec k,\sigma,\tau,\alpha} U_{\alpha}
(f^{\dag}_{\sigma} b_{\bar{\tau}} 
c^{\alpha}_{\vec k\sigma\tau} + h.c.),
\label{hamiltonian} 
\end{eqnarray}
where $f_{\sigma}$, ($\sigma = 1,\dots N$) transforms according to SU(N) and
$b_{\bar{\tau}}$, ($\bar\tau =1,\dots M$) transforms according to the
adjoint representation of SU(M). 
The first term in Eq.~(\ref{hamiltonian})
describes the conduction electron bands with kinetic energy 
$\varepsilon _{\vec p}$ offset by $-eV_{\alpha}$ due to 
an applied voltage. The index $\alpha$ is equal to $L$, $R$
for the left and the right reservoirs, respectively.
Note that the two reservoirs do not constitute different scattering 
channels in the sense of the multi--channel model, since
the reservoir index $\alpha$ is {\it not} conserved by the Kondo
interaction. The second and third terms represent the energy of d--states 
and the hybridization term, respectively.  The constraint term is
not explicitly written in the Hamiltonian Eq.~(\ref{hamiltonian}). Note that
the local charge $Q$ commutes with the Hamiltonian.

As discussed in the introduction, there are a number of possible
physical realizations of the one-channel non--equilibrium Kondo model:
magnetic impurities in tunnel junctions,
tunneling through charge
traps, and possibly even tunneling through quantum dots.
In each of these models the d-states introduced in the Anderson
model have physical meaning.  In the case of a transition
metal magnetic impurities
the d-states are literally the atomic d-states of the impurity.
For a charge trap, the d-states are the electronic states for
the two possible spin orientations of the trap.
The two-channel model  has been proposed as a possible 
scenario for the occurrence of non--Fermi liquid behavior in some 
heavy fermion compounds with cubic crystal symmetry.\cite{coxzawa,cox2,cox4}
In that case the occupied d--states correspond to the states 
of a low--lying non-magnetic doublet of the rare earth or actinide atoms,
while the empty levels, described by $b_{\bar{\tau}}$, constitute an
excited doublet of local orbitals\cite{cox4}.

On the other hand, for
the physical realization in terms of two-level-systems,
the  empty states ($b_{\bar{\tau}}^{\dag}$) 
do not have direct physical meaning.  They
are introduced as a construct in representing the pseudo--spins such that
the channel quantum number $\tau$ is conserved by the Kondo interaction.
Via a Schrieffer--Wolff transformation \cite{schriwo}
one can show that the low energy physics of the
Anderson model of Eq. (\ref{hamiltonian}) is
the same as for the Kondo model in the limit when the occupation of
the d-states $n_d$ approaches one. Thus, although we use the Anderson
model, the results for the low energy physics
are expected to be the same as for the Kondo model.

\subsection{The Non--crossing Approximation (NCA)}
\subsubsection{Validity of the NCA}

In the present context we are interested in the Kondo regime of the Anderson
model Eq.~(\ref{hamiltonian}), where the low energy effective coupling 
$J_{\alpha}=|U_{\alpha}|^2/\varepsilon _d$ between the band electrons and the
impurity is small, ${\cal N}(0)J_{\alpha}\ll 1$, with ${\cal N}(0)$ the 
band electron density
of states per (pseudo--) spin and channel. The NCA is a self--consistent
conserving perturbation expansion for the pseudo--fermion and slave 
boson self--energies
to first order in ${\cal N}(0)J_{\alpha}$. Considering the inverse level 
degeneracy $1/N$ as an expansion parameter, the NCA includes all 
self--energy diagrams up to ${\cal O}(1/N)$. The self--energies are then made 
self--consistent by inserting the
dressed slave particle propagators in the Feynman diagrams instead of the bare
propagators.\cite{cole,kuram,muhart,bickers}
It is easiliy seen that this amounts to the summation of all 
self-energy diagrams without any  
propagator lines crossing each other, hence the
name Non-crossing Approximation.    

One may expect that the self--consistent perturbative approach is valid
as long as the summation of higher orders in $J_{\alpha}$ or $1/N$ do not 
produce additional singularities of perturbation theory. 
It has recently been shown\cite{kroha2,kroha,saso} that such
a singularity does arise in the single--channel case ($M=1$, $N=2$) below the
Kondo temperature $T_K$ due to the
incipient formation of the singlet bound state between conduction electrons and
the local impurity spin. However, around and 
above $T_K$ and in the Kondo limit of the 
two--channel model ($N=2$, $M=2$) even down to the lowest temperatures  
this singularity is not present\cite{kroha2}. Indeed, the NCA has
been very successful in describing the single--channel Kondo model except for
the appearance of spurious non--analytic behavior at temperatures far below 
$T_K$. The spurious low temperature properties are due to the fact that the  
NCA neglects vertex corrections responsible for restoring the Fermi liquid 
behavior of the single--channel model.\cite{costi,kroha2}. 
A qualitatively correct description was 
achieved\cite{bickers,costi} for the wide temperature range from well below $T_K$
(but above the breakdown temperature of NCA) through the crossover region around
$T_K$ up to the high temperature regime $T > T_K$.
For the multi--channel problem, $M\geq N$, the complications of the
appearance of a spin screened Fermi liquid fixed point are absent.
For this case it has recently been shown\cite{cox3} that the NCA
does reproduce the exact\cite{bethemulti,afflud} 
low--frequency power law behavior of all physical properties
involving the 4--point slave particle
correlation functions, like the impurity spectral function $A_d$ and the
susceptibilities, down zero temperature. Therefore, in the multi--channel case
the NCA is a reliable approximation\cite{taesuk} 
for quantities involving $A_d$ (like the 
non--equilibrium conductance) and the susceptibilities, even at the 
lowest temperatures. 

\subsubsection{NCA in thermodynamic equilibrium}
The slave boson perturbation expansion is initially formulated
in the grand canonical ensemble, i.e.~ in the enlarged Hilbert space of
pseudo--fermion and slave boson degrees of freedom, with a single chemical
potential $-\lambda$ for both pseudo--fermions and slave bosons. Therefore,  
standard diagram techniques are valid, including Wick's theorem. 
In a second step, the exact projection
of the equations onto the physical Hilbert space, $Q=1$, is 
performed\cite{cole,bickers,mewin3}. For a brief review of the projection
technique we refer the reader to appendix A.

The equations for the self--energies of the retarded Green functions 
of the pseudo-fermions,
$G^r(\omega) = (\omega -\epsilon_d - \Sigma^r (\omega))^{-1}$,
and the slave-bosons, $D^r(\omega) = (\omega  - \Pi^r (\omega))^{-1}$,
constrained to the physical subspace, read
\begin{mathletters}
\begin{eqnarray}
\Sigma^r(\omega) &=& M\frac{\Gamma}{\pi} \int d\epsilon 
\bar{\cal N} (\omega -\epsilon)
f(\epsilon -\omega ) D^r(\epsilon) \\
\Pi^r(\omega) &=& N\frac{\Gamma}{\pi} \int d\epsilon 
\bar{\cal N} (\epsilon - \omega)
f(\epsilon - \omega ) G^r(\epsilon) \,\,\, ,
\end{eqnarray}
\label{ncaraw}
\end{mathletters}

\noindent
where $\Gamma = \pi |U|^2 {\cal N}(0)$,
$\bar{\cal N}(\omega)={\cal N}(\omega)/{\cal N}(0)$ is the bare density
of states of the band electrons, normalized to its value at the Fermi level, 
and $f(\omega )=1/(1+e^{\beta \omega})$ is the Fermi function.
The real and imaginary parts of the self-energy
are related via Kramers--Kroenig relations, e.g.
\begin{equation}
\mbox{Re}\Sigma^r(\omega) = \frac{1}{\pi} {\cal P} \int d\epsilon \frac{\mbox{Im}
\Sigma^r(\epsilon)} {\epsilon - \omega}. 
\label{krakro}
\end{equation}
Taking the imaginary part of Eqs. (\ref{ncaraw}) 
and defining the spectral functions for the slave particles as, 
\begin{eqnarray}
A(\omega )&=& -\mbox{Im}G^r(\omega )/\pi = -\mbox{Im}\Sigma^r(\omega)\,
|G^r(\omega)|^2/\pi \nonumber\\
B(\omega )&=& -\mbox{Im}D^r(\omega )/\pi = 
-\mbox{Im}\Pi^r(\omega)\,|D^r(\omega)|^2/\pi \nonumber\\
\end{eqnarray}
we arrive at the self--consistent equations 
\begin{mathletters}
\begin{eqnarray}
\frac{A(\omega)}{|G^{r}(\omega)|^{2}} &=& 
M\frac{\Gamma}{\pi} \int d\epsilon 
\bar{\cal N} (\omega -\epsilon) f(\epsilon -\omega) B (\epsilon) \\ 
\frac{B(\omega)}{|G^{r}(\omega)|^{2}} &=& 
N\frac{\Gamma}{\pi} \int d\epsilon 
\bar{\cal N} (\epsilon - \omega) f(\epsilon - \omega ) A (\epsilon) \, .
\end{eqnarray}
\label{ncareteq}
\end{mathletters}

\noindent
Together with the Kramers-Kroenig relations, 
Eq. (\ref{krakro}),  
Eqs. (\ref{ncareteq}) form a complete set of equations 
to determine the slave particle propagators.
However, an additional difficulty arises in the construction of physical
quantities from the auxiliary particle propagators. 
The local impurity propagator, $G_{d,\sigma\tau}(\tau -\tau ') 
= -\langle \hat T\{d_{d,\sigma\tau}(\tau )d_{d,\sigma\tau}^{\dag}(\tau ')\}\rangle$,
is given by the $f-b$ correlation function. Thus, its spectral function is
calculated within NCA as\cite{muhart}
\begin{eqnarray}
A_{d}(\omega) = \frac{1}{Z} \int d\epsilon e^{-\beta\epsilon}
\left[\, A(\epsilon + \omega) B(\epsilon) + A(\epsilon) B(\epsilon -\omega) 
\right],
\label{spectral}
\end{eqnarray}
where  
\begin{eqnarray}
Z = \int d\epsilon e^{-\beta\epsilon} [ N A(\epsilon) + M B(\epsilon) ]
\label{partition}
\end{eqnarray}
is the canonical partition function of the impurity in the physical Hilbert space,
$Q=1$ (see appendix A). 
The requirement that $Z$ be finite implies that the auxiliary particle
spectral functions vanish exponentially below a threshold energy
$E_o$.\cite{kuram,muhart}
Above the threshold, the spectral functions show characteristic power law
behavior originating\cite{kroha2} from the Anderson orthogonality catastrophe. 
In Eqs.~(\ref{spectral}), (\ref{partition}) the Boltzmann factor does not
allow for a direct numerical evaluation of the integrand at negative $\epsilon$ 
if $\beta = 1/k_B T$ is large. It is therefore necessary to absorb the
Boltzmann factor in the spectral functions and find solutions for the 
functions 
\begin{equation}
a(\omega) = e^{-\beta\omega} A(\omega)\,\, , \,\,\,\,\,
b(\omega) = e^{-\beta\omega} B(\omega)\,\, . 
\label{retles}
\end{equation}
Using $e^{\beta\omega} f(\omega)=f(-\omega)$, 
the equations determining $a(\omega )$ and $b(\omega )$ are easily found
from Eqs. (\ref{ncareteq}):
\begin{mathletters}
\begin{eqnarray}
\frac{a(\omega)}{|G^{r}(\omega)|^{2}} = 
M\frac{\Gamma}{\pi} \int d\epsilon 
\bar{\cal N}(\omega -\epsilon) f(\omega -\epsilon) b (\epsilon) \\
\frac{b(\omega)}{|G^{r}(\omega)|^{2}} = 
N\frac{\Gamma}{\pi} \int d\epsilon 
\bar{\cal N}(\epsilon - \omega) f(\omega -\epsilon) a (\epsilon) \,\,\, .
\end{eqnarray}
\label{ncaleseq}
\end{mathletters}

\noindent
The equations for the impurity
spectral function  and the partition function then become 
\begin{eqnarray}
A_d(\omega) = \frac{1}{Z} \int d\epsilon 
\left[\, A(\epsilon + \omega) b(\epsilon) + a(\epsilon) B(\epsilon -\omega) 
\right]   \label{specfun} 
\end{eqnarray}
\begin{eqnarray}
Z = \int d\epsilon \left[ N a(\epsilon) + M b(\epsilon) \right] \, .
\label{partfun}
\end{eqnarray}
In view of the generalization to non--equilibrium, 
it is instructive to realize that the functions $a(\omega)$ and $b(\omega)$ 
are proportional to the Fourier transform of the lesser Green functions used in the
Keldysh technique\cite{keldysh},
\begin{eqnarray} 
a(\omega) &=& \frac{i}{2\pi} G^< (\omega) , \,\,\,\,\, 
G^{<}(t-t')= -i\langle\, f^{\dag}(t') f(t)\,\rangle\nonumber\\
b(\omega) &=& \frac{i}{2\pi} D^< (\omega), \,\,\,\,\, 
D^{<}(t-t')= i\langle\, b^{\dag}(t') b(t) \rangle ,
\end{eqnarray}
and contain information about the distribution functions of the slave particles.
Henceforth we will call $a(\omega)$ and $b(\omega)$ the `lesser' functions.
Eqs.~(\ref{ncareteq}), (\ref{ncaleseq}) and (\ref{krakro}) form a set of 
self--consistent
equations which allow for the construction of the impurity spectral function
$A_d$.\\ 

A significant simplification of the above procedure can be achieved by
exploiting that {\it in equilibrium} the Eqs.~(\ref{ncareteq}) and (\ref{ncaleseq})
are not independent but linked to each other by Eq.~(\ref{retles}). 
Hence, we define
new functions $\tilde A (\omega)$ and $\tilde B (\omega)$ via\cite{kroha}
\begin{eqnarray}
f(-\omega)\tilde A (\omega) = A(\omega)\,\, , \,\,\,\,
f(-\omega)\tilde B (\omega) = B(\omega).
\label{tilfun}
\end{eqnarray}
By definition, $\tilde A (\omega)$ and $\tilde B (\omega)$ do not have
threshold behavior, and the spectral functions as well as the lesser functions
may easily be extracted from them, i.~e.~$a(\omega)=f(\omega)\tilde A(\omega)$,
$B(\omega)=f(\omega)\tilde B(\omega)$. Inserting Eq.~(\ref{tilfun}) into 
Eqs.~(\ref{ncareteq}) one obtains the NCA equations for 
$\tilde A (\omega)$ and $\tilde B (\omega)$,
\begin{mathletters}
\begin{eqnarray}
\frac{\tilde A (\omega)}{|G^{r}(\omega)|^{2}} &=& 
M\frac{\Gamma}{\pi} \int d\epsilon 
\bar{\cal N}(\omega -\epsilon) \frac{f(\epsilon -\omega)) f(-\epsilon)}
{f(-\omega)}
\tilde B (\epsilon) \\
\frac{\tilde B (\omega)}{|G^{r}(\omega)|^{2}} &=& 
N\frac{\Gamma}{\pi} \int d\epsilon 
\bar{\cal N}(\epsilon -\omega) \frac{f(\epsilon -\omega) f(-\epsilon)}
{f(-\omega)} \tilde A (\epsilon).
\end{eqnarray}
\label{ncatil}
\end{mathletters}

\noindent
One can convince oneself that the statistical factors appearing in
these equations  
are non--divergent in the zero temperature limit for all frequencies
$\omega$, $\epsilon$. Thus, by solving the two Eqs.~(\ref{ncatil}) instead of
the four Eqs.~(\ref{ncareteq}), (\ref{ncaleseq}), one saves a significant
amount of integrations.
The equations are solved numerically by iteration.  
After finding the solution at an elevated temperature, $T$ is gradually
decreased. As the starting point of the iterations at any given $T$ we take 
the solution at the respective previous temperature value.  In appendix
A we describe an elegant and efficient implementation of the NCA equations
which leads to a significant improvement in computational precision as well
as speed. The proper setup of the discrete frequency meshes for the 
numerical integrations in the equilibrium and in the non--equilibrium case
is dicussed in some detail in appendix B. In this way
temperatures of 1/1000 $T_K$ and below may be reached without much
effort. The solutions we obtained fulfill the exact sum rules 
\begin{eqnarray}
n_d \equiv N\int &d&\epsilon \, f(\epsilon) A_d (\epsilon) =
N{\int }d\epsilon  \, a(\epsilon) \equiv n_f \nonumber\\
\int &d&\epsilon  \, A_d (\epsilon) = 1 - (1-\frac{1}{N}) n_f \nonumber
\end{eqnarray}
typically to within 0.1\% or better, where $n_d$ and $n_f$ are the occupation
numbers of physical $d$--particles and pseudo--fermions in the impurity level,
respectively.\\   

An important quantity is the self--energy $\Sigma _c(\omega)$
of the conduction electrons due to scattering off the Kondo or Anderson 
impurities. In the limit of dilute impurity concentration $x\ll 1$,
it is proportional to the bulk (linear response) 
resistivity of the system and determines 
the renormalized conduction electron density of states, which can be
measured in tunneling experiments. Below we will calculate $\Sigma _c(\omega)$
within NCA in order to compare with the CFT prediction for the resistivity
in equilibrium on one hand, and to compare the linear response result with
the zero bias conductance calculated from a generalized Landauer--B\"uttiger
formalism (see section III A) on the other hand.

$\Sigma _c(\omega)$ is defined via the impurity averaged
conduction electron Green function in momentum space, 
$G_{c\, {\bf k}}(\omega )=[\omega -\varepsilon _{\bf k}-\Sigma _c(\omega)]^{-1}$. 
In the dilute limit and for pure s--wave scattering, 
$\Sigma _c(\omega)$ is momentum independent,
\begin{eqnarray}
\Sigma _c(\omega)= xt(\omega ) \,
\label{sigceq}
\end{eqnarray}
where $t(\omega )$ is the local $T$--matrix for scattering off a single impurity.
According to the Hamiltonian, Eq.~(\ref{hamiltonian}),
$t(\omega )$ is given exactly in terms the local $d$--particle propagator
and reads, e.g.~for scattering across the junction ($L\rightarrow R$),
\begin{eqnarray}
t(\omega ) = U_R U_L^* G_d(\omega )\, .
\label{tmatrix}
\end{eqnarray}

\subsection{NCA for static non--equilibrium}
If we apply a finite bias $V$, 
the system is no longer in equilibrium. 
We cannot expect the simple relation Eq. (\ref{retles})
between the lesser and the spectral functions to hold in this case.
Therefore, the trick
with introducing the functions $\tilde A$ and $\tilde B$ cannot be performed.
Rather, the NCA equations have to be derived by means of standard 
non--equilibrium Green function techniques,\cite{mewin3,keldysh,langreth}    
and one has to solve the equivalent of Eqs.~(\ref{ncareteq}) and (\ref{ncaleseq})
for the non--equilibrium case without any further simplification.
Defining in analogy to the equilibrium case
$\Gamma_{L,R}  = \pi |U_{L,R}|^2 {\cal N}(0)$,
the NCA equations for steady state non--equilibrium are
\begin{mathletters}
\begin{eqnarray}
&&\frac{A(\omega)}{|G^{r}(\omega)|^{2}} = 
\frac{M}{\pi} \int d\epsilon B (\epsilon) \nonumber \\
&&\times \sum_{\alpha =L,R} \left[ \Gamma_{\alpha}\bar{\cal N}
(\omega -\epsilon + \mu_{\alpha}) 
f(\epsilon -\omega -\mu_{\alpha})\right]  \\ 
&&\frac{B(\omega)}{|G^{r}(\omega)|^{2}} =
\frac{N}{\pi} \int d\epsilon A (\epsilon) \nonumber \\
&&\times \sum_{\alpha =L,R} \left[ 
\Gamma_{\alpha} \bar{\cal N}
(\epsilon - \omega - \mu_{\alpha}) f(\epsilon - \omega 
-\mu_{\alpha})\right] 
\end{eqnarray}
\label{ncaret}
\end{mathletters}
\begin{mathletters}
\begin{eqnarray}
&&\frac{a(\omega)}{|G^{r}(\omega)|^{2}} = 
\frac{M}{\pi} \int d\epsilon \,b (\epsilon)\nonumber \\
&&\times \sum_{\alpha =L,R} \left[ \Gamma_{\alpha} \bar{\cal N}(\omega -\epsilon + \mu_{\alpha}) 
f(\omega -\epsilon +\mu_{\alpha}) \right] \\
&&\frac{b(\omega)}{|G^{r}(\omega)|^{2}} = 
\frac{N}{\pi} \int d\epsilon \,a (\epsilon) \nonumber \\
&&\times \sum_{\alpha =L,R} \left[ \Gamma_{\alpha} \bar{\cal N}(\epsilon - \omega - \mu_{\alpha}) 
f(\omega -\epsilon + \mu_{\alpha})\right]  .
\end{eqnarray}
\label{ncales}
\end{mathletters}

\noindent
If the density of states ${\cal N}(\omega)$ were a constant, the only difference
between the equilibrium and the non--equilibrium NCA equations would be 
the replacement of the Fermi function by an
effective distribution function $F_{eff}$ given by 
\begin{equation}
F_{eff}(\epsilon) = \frac{\Gamma_L}{\Gamma_{tot}} f(\epsilon
-\mu_L) +  \frac{\Gamma_R}{\Gamma_{tot}} f(\epsilon -\mu_R) ,
\end{equation}
where $\Gamma_{tot} = \Gamma_L + \Gamma_R$.
Since our density of states is a Gaussian with a width much larger than all the 
other energy scales, $|\epsilon_d |,\, \Gamma_{tot},\, T_K$, 
this is in fact
the only significant modification of the NCA equations.
Numerically, the most crucial modification concerns the integration mesh.
The proper choice of  integration meshes is central to the success of the
iteration and is discussed in Appendix B.

\section{Current formulae, conductance and susceptibilities}

\subsection{Current formulae and conductance}
For the case of tunneling through a Kondo impurity, 
the current is directly related to the impurity Green functions.
In particular, the current in the left or in the right lead is
given\cite{sdw,mewin2} by a generalized Landauer--B\"uttiger formula,
\begin{mathletters}
\begin{eqnarray}
I_L(V) &=& -N\frac{e}{\hbar}\Gamma_L
\int d\omega\bar{\cal N}(\omega -\mu_L)\nonumber \\
&\times &\left[ G_d^< (\omega) -   A_{d}(\omega) f(\omega - \mu_L)\right]\\
I_R(V) &=& N \frac{e}{\hbar}\Gamma_R
\int d\omega\bar{\cal N}(\omega -\mu_R)\nonumber \\ 
&\times &\left[ G_d^< (\omega) -   A_{d}(\omega) f(\omega - \mu_R)\right],
\label{lericurr}
\end{eqnarray}
\end{mathletters}

\noindent
where $G_d^<$ is the lesser Green function of the impurity.
It is obtained from the pseudo--fermion and slave boson Green functions via
\begin{equation}
G_d^< (\omega) = \frac{1}{Z}\int d\epsilon\, 
a(\epsilon) B(\epsilon -\omega)\, .
\end{equation}
Making use of current conservation, $I_L = I_R$, and taking the 
wide band limit, where  ${\cal N}(\omega )$ is taken to be a constant,
the current may be expressed solely in terms of the impurity
spectral function
\begin{eqnarray}
I(V) &=& N\frac{e}{\hbar}
\frac{2\Gamma _L\Gamma _R}{\Gamma _L + \Gamma _R}
\int d\omega A_{d}(\omega) 
\nonumber \\
&\times & [f(\omega -\mu_L) - f(\omega - \mu_R)].
\label{curr}
\end{eqnarray}
The NCA is a conserving approximation.\cite{mewin3} 
Therefore, the currents
computed for the left and the right leads should be the same when 
evaluated numerically.
We have checked the current conservation within NCA and 
found that the two currents agree to within 0.5\%, which
sets a limit to the uncertainty for the average current,
$I(V) = (I_L + I_R)/2$.

In order to obtain the differential conductance,
$G(V)= dI(V)/dV$, we perform the numerical derivative 
$(I(V_1) - I(V_2))/(V_1 - V_2)$, and take it as the value of $G(V)$ 
$V=(V_1 + V_2)/2$. The numerical error involved in this procedure 
could be reduced to as little as 2\%.
The zero bias conductance (ZBC) is the special case of the above
equations in the limit of vanishing applied voltage $V\rightarrow 0$.
The ZBC for a tunnel junction is thus
\begin{equation}
G(0,T) = N \frac{e^2}{\hbar}
\frac{2\Gamma _L\Gamma _R}{\Gamma _L + \Gamma _R}
 \int d\omega \left(-\frac{\partial f(\omega)}{\partial \omega}\right) 
A_d(\omega) .
\label{ZBCtun}
\end{equation}
It will be useful to compare this to the linear response bulk
resistivity for a small density of impurities in a metal.
The resistivity, $\rho$,
is related to the impurity spectral function via \cite{bickers}
\begin{equation}
1/\rho = const \int d\omega \left(-\frac{\partial f(\omega)}
{\partial \omega}\right) \tau (\omega),
\end{equation}
where the impurity scattering rate is
$\tau^{-1}(\omega) = x U_L U_R^* A_d(\omega)$.  The impurity concentration is 
denoted by $x$.

Most of our calculations were done with symmetric couplings, 
$\Gamma_L=\Gamma_R$. However, this is not necessarily the
case in an experimental situation, especially for tunnel junctions. When 
an Anderson impurity is placed inside a tunneling barrier of thickness $d$, the 
tunneling matrix element $U_{\alpha}$
depends exponentially on the distance $z$ of the impurity from the surface of the 
the barrier. Also, the bare energy level
$\varepsilon_d$ of the impurity will be shifted due to the approximately 
linear in $z$ voltage drop inside the barrier.
In order to investigate the consequences on the non--equilibrium
conductance, we also performed evaluations with asymmetric couplings. 
For simplcity, and in order to keep the total coupling $\Gamma_{tot} 
= \Gamma_L + \Gamma_R$ constant, 
we assume a linear dependence of the 
$\Gamma_{\alpha}$'s on $z$ of the form $\Gamma_L = \Gamma_{tot} (1 - z/d)\, ,
\,\,\, \Gamma_R = \Gamma_{tot}  z/d$. We also modify  
$\epsilon_d$ according to
$\epsilon_d (V) = \epsilon_d + (V/2) (1 - 2z/d)$. The latter modification
turns out to be insignificant as long as $V  |\epsilon_d|$.

\subsection{Tunnel junctions vs. point contacts}
The above formulae for the currents and conductances
are valid in a tunnel junction geometry
where the current must flow through the impurity. In a point contact the
two leads are joined by a small constriction.  A current $I_o$
will flow through
the constriction without the impurity being present. In fact, the impurity 
will \sl impede \rm the current due to additional scattering in the vicinity 
of the constriction. 
The question arises whether the effect of an impurity
in a point contact is the same in magnitude but opposite in sign.
In Appendix C we derive a general formula for the conductance which
allows one to go continuously between a clean point contact and 
a tunnel junction.  In the limit of a clean point contact, where 
the transmission probabilities are close to unity, we find that
the change in the conductance due to an impurity in a point contact has
the same form as for a tunnel junction, except for a change in sign.
Thus, in clean samples 
the results for the current calculated for the tunnel junction apply for 
point contacts as well, if one subtracts out the background current, $I_o$.
If $I_o$ is ohmic, the conductance $G(V)$ is shifted
by the constant $dI_o/dV$. Aside from this shift and sign difference, 
the conductance signals
of a tunnel junction and a clean point contact will be the same. 

\subsection{Susceptibilities}
The impurity contribution to the dynamic \hbox{(pseudo-)} spin
susceptibility is calculated using the standard formulae
\cite{muhart,bickers} from the lesser and the spectral function of the 
pseudo-fermions. The formula for the imaginary part reads
\begin{equation}
\mbox{Im}\chi(\omega) = \frac{1}{Z} \int \frac{d\epsilon}{\pi} 
\left[\, A(\epsilon + \omega) a(\epsilon) - a(\epsilon) A(\epsilon -\omega)
\right]\,\,\, .
\label{suscep}
\end{equation}
The real part can be obtained by means of a Kramers--Kroenig relation: 
\begin{equation}
\mbox{Re}\chi(\omega) =  \frac{1}{\pi} {\cal P} \int d\epsilon \frac{\mbox{Im} 
\chi(\epsilon)} {\epsilon -\omega} .
\end{equation}
The static susceptibility $\chi_o = \chi (\omega = 0)$ follows directly from 
this equation. 
Note that in the two--channel Anderson model as possibly realized in TLS's,
this susceptibility is not
the magnetic susceptibility. Rather, it is probed by a field coupling to the
impurity pseudospin, e.g.~a crystal field breaking the degeneracy of the TLS.

\section{Results}
\subsection{Conductance for one-- and two--channel models 
with symmetric couplings}
Using the formulae discussed in the previous section,
we now present the results obtained from the numerical
evaluation of the bulk resistivity and of the
conductance for symmetric couplings. 
For the evaluations a 
Gaussian conduction electron density of states ${\cal N}(\omega)$
with half width $D$ was used. All calculations were done in the Kondo regime 
for the set of 
parameters $\varepsilon _d=-0.67D$, $\Gamma_L=\Gamma_R=0.15D$.
In order to make the most direct comparison to experiment,
the results for the two--channel case 
have been computed for a \underline{point contact},
and the results for the one
channel case have been computed for 
a \underline{tunnel junction}, except for Fig. \ref{scal12}
where we compare the scaling behavior of the nonlinear conductance
for the one-- and two--channel models.

\subsubsection{Linear response conductance and resistivity}
The low temperature limit of the linear response conductance
shows power law behavior in temperature.  
The exponent is determined by the symmetry of the underlying Kondo model.
As explained in the discussion of the NCA, we expect to 
get quantitatively correct behavior for the two channel model,
but not for the one channel case.
In Fig.~\ref{covst2ch} we show the zero bias correction to the
conductance  $G(0,T)$
for a two channel Kondo impurity ($N=M=2$) in a point contact.
The zero bias conductance does show the expected\cite{afflud} $T^{1/2}$
dependence at low $T$. 
Deviations from this power law start at about
1/4 $T_K$, where the Kondo temperature $T_K$ is determined from the data as
the width at half maximum of the zero bias impurity spectral function, $A_d$,
at the lowest calculated $T$ (see Fig. \ref{adfig}).
The slope of the $T^{1/2}$ behavior defines a constant
$B_\Sigma$:
\begin{figure*}[t]
\leavevmode\centering\psfig{file=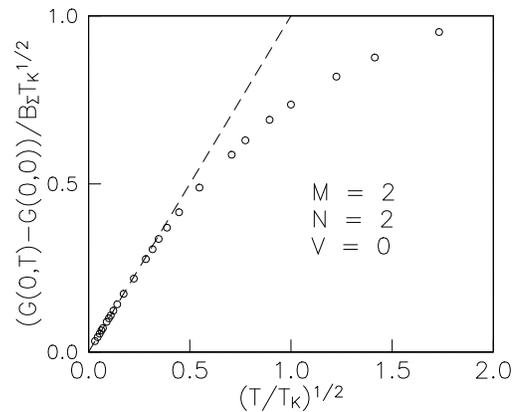,width=2.7 in}
\caption{Temperature dependence of the zero bias conductance 
for the two--channel model ($M=2$, $N=2$) in a point contact.
The zero bias conductance has $T^{1/2}$ dependence for $T< T_K/4$. 
This can be used to roughly extract $T_{K}$ from the experimental data.
$B_{\Sigma}$ (compare Eq.~(27))
is a material dependent constant which has been divided out.
Therefore, the slope of the low $T$ fit (dashed line) is equal to unity.}
\label{covst2ch}
\end{figure*}
\begin{equation}
G(0,T) - G(0,0) =  B_{\Sigma} T^{1/2} ,
\end{equation}
which we will use below in interpreting the nonlinear conductance.

\begin{figure*}[t]
\leavevmode\centering\psfig{file=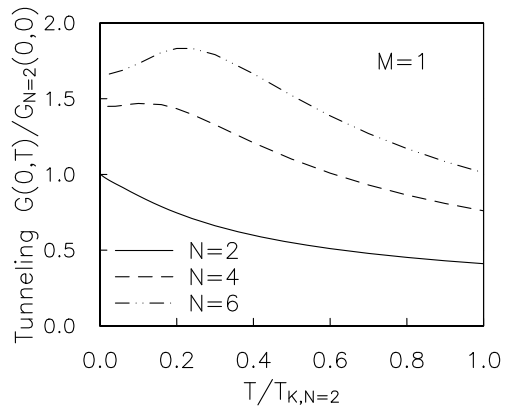,width=2.7 in}
\caption{Zero bias conductance for  tunneling through a single--channel
Anderson impurity ($M=1$, $N=2$) vs. temperature. 
The conductance for a  clean point contact in the presence of a single--channel
Kondi impurity would 
be obtained by subtraction of this curve from a (constant) background 
conductance.
The graph for $N = 2$ shows an almost linear $T$ dependence at low $T$
whereas the curves for
spin degeneracy $N = 4$ and $N=6$ show non--monotonic behavior.
The humps are due to the fact that the Kondo peak of the spectral function 
$A_d(\omega)$ is shifted away from the Fermi energy $\epsilon_F$ by about $T_K$.
For $T > T_K$ all the curves fall like log$(T/T_K)$ for approximately 
one decade.}
\label{covst1ch}
\end{figure*}

On the other hand,
for the one channel case ($M=1, N=2$) one  
expects $T^2$ dependence because of the Fermi liquid behavior at low 
temperatures.
As shown in Fig. \ref{covst1ch} for a tunnel junction, the NCA as a 
large N expansion is not able to reproduce this power law for $N=2$ at temperatures
below $T_K$. Increasing $N$ to
$N=4$ and $N=6$, the ZBC develops a hump as a function of 
\noindent temperature.
This peak is due to the fact that the Kondo resonance is shifted away
from the Fermi level for $N>2$. 
Although we know of no experimental evidence for such humps in
zero bias anomalies, similar humps have be seen in the magnetic
susceptibilities of these systems.\cite{bickers}
Note that for $N=6$, a $T^2$ behavior seems to appear 
at temperatures below the hump. The temperature range shown here
is above the breakdown temperature of NCA, below which a fractional power law
$G(0,T)-G(0.0)\propto - T^{M/(M+N)}$ would appear. 

\begin{figure*}[b]
\leavevmode\centering\psfig{file=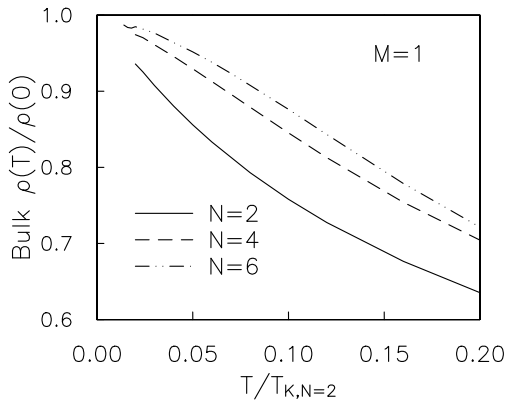,width=2.7 in}
\caption{Bulk resistivity vs. temperature for the $M=1$ channel model, $N=2,4,6$.
Of the three curves only $N = 6$ has a clear convex shape 
and falls roughly like $T^2$ at low $T$. The $N = 2$ graph again shows almost 
linear $T$ dependence. Note that the humps in the conductance for 
$N=4$ and $N=6$ are not present in the bulk resistivity $\rho$.}
\label{revst1ch}
\end{figure*}

For a bulk Kondo system it
is impossible to measure the zero bias conductance
of single impurities. Instead, one 
measures the linear response resistivity, $\rho$.
In Fig. \ref{revst1ch} we show the impurity contribution to
the resistivity for one channel impurities with $N=2,4,6$.
Only the $N=6$ curve shows a convex dependence on $T$. In fact, $\rho$ seems
to behave like
$(1 - const (T/T_K)^2)$ at the temperatures shown, consistent with a Fermi liquid.
\cite{bickers}  For $N=2$ there is no convex temperature dependence even 
down to $T = 0.02 T_K$.
Figures \ref{covst1ch} and \ref{revst1ch} also serve to illustrate
that the zero bias conductance and the bulk resistivity for the 
same kind of Kondo impurities do not necessarily have the same
temperature dependence.
\subsubsection{Nonlinear conductance}
Recently, it has been shown \cite{hkh1} that the two--channel model exhibits
scaling of the nonlinear conductance $G(V,T)$ as a function of bias $V$
and $T$ of the form \cite{rlvdb}
\begin{equation}
G(V,T) - G(0,T) =  B_{\Sigma} T^{\eta} H( A\frac{eV}{k_{B}T}) \,\, .
\label{scal}
\end{equation}
Here, $H$ is a universal scaling function which satisfies
$H(0) = 0$ and $H(x) \propto x^{\eta}$ for $x\gg 1$, and $A$, $B_{\Sigma}$ 
are non-universal constants. 
The exponent $\eta$ is 1/2 for the two--channel model.
This scaling ansatz is motivated by 
\begin{figure*}[h]
\leavevmode\centering\psfig{file=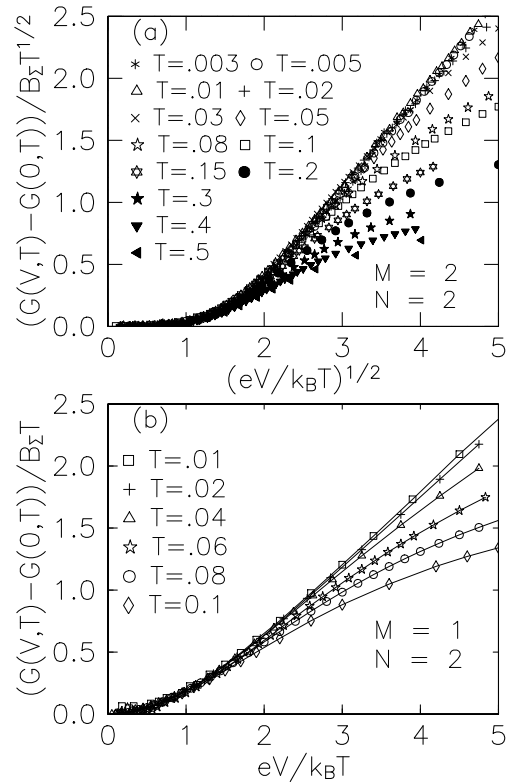,width=2.7 in}
\caption{Scaling plots of the conductance of point contacts in the
presence of (a) a two--channel impurity 
($M=N=2$) and (b) a one--channel impurity ($M=1$, $N=2$).
With $\Gamma _L = \Gamma _R$ and 
$B_\Sigma$ determined from the zero bias conductance (compare Eq.~(27)),
there are no adjustable parameters. 
There are two regimes in these plots.
For $(eV/k_BT)^{\eta} < 1.5$ the curves collapse onto a single
curve and the rescaled conductance is proportional to
$(eV/k_BT)^2$.
For larger $(eV/k_BT)^{\eta}$ the rescaled conductance 
is linear on these plots. There are substantial 
corrections to scaling even at $T$ small compared to $T_K$.
At even larger biases this linear behavior rounds off, indicating 
the breakdown of scaling.
The temperatures are given in units of the respective $T_K$ for the two-- and the
one--channel case.}
\label{scal12}
\end{figure*}
\noindent
the scaling of the conduction electron 
self--energy in the variables frequency $\omega$ and temperature $T$ 
as obtained by CFT in equilibrium.\cite{afflud}  
Scaling behavior is well known\cite{wilson} to 
be present also in the equilibrium properties of the single--channel model 
($M=1$, $N=2$). Hence, in the case $M=1$ one may expect a scaling form of the 
non--equilibrium conductance similar to Eq. (\ref{scal}) as well,
however with Fermi liquid exponent $\eta = 2$.

In order to examine whether the scaling ansatz is correct, 
the rescaled conductance is plotted as a function of
$(eV/k_{B}T)^{\eta}$. The conductance curves for  different  $T$ should 
collapse onto a single curve with a linear part for not too large and not too
small arguments: Very large $V$ or $T$ would drive the system out of the scaling
regime. A  collapse indeed occurs for low bias $V < T$. However,
for larger bias the slope of the linear part shows $T$ dependence
(see Ref.~[\cite{hkh1}] for more details).
This shows that there are significant $T$--dependent
corrections to scaling, indicating that finite bias $V$ and finite temperature
$T$ are not equivalent as far as scaling is concerned, although both parameters
have qualitatively similar effects on the conductance. 

Figs. \ref{scal12} (a) and (b) show the scaling plots for the cases 
$M=2$ and $M=1$,
respectively, with $N=2$ in both cases. Whereas the two--channel case 
shows the behavior described above with the expected exponent $\eta = 1/2$, the
NCA does not give the correct exponent for the single--channel model.
In fact, the data show approximate scaling, however, the 
exponent $\eta$ extracted from the NCA data appears to be equal 
to unity rather than 2. This seems to reflect the dominant linear
temperature dependence of the ZBC that the NCA produces in this case.
This shortcoming is another consequence of the negligence of singular 
vertex corrections within the NCA. 

\subsection{Self--energy scaling and comparison to  CFT}
Two possible origins for the above mentioned finite--$T$ corrections 
to scaling at low temperatures $T$ are: (i) The
non--equilibrium state brings about terms in the 
electron self-energy which break scaling of the form given by CFT, and (ii)
the deviations from scaling in equilibrium at finite $T$ 
have large coefficients, restricting the scaling regime to $T \ll T_K$.  

Recently, it has been shown for the single--channel 
model\cite{schill-hersh} that in non--equilibrium the conductance
indeed has terms which explicitly break the scaling behavior.
Though the coefficients of these terms are small, 
scaling in the ordinary sense is clearly violated, i.e. at 
temperatures  well below a crossover temperature ($T_K$) 
not all corrections to scaling vanish.
It is quite possible that the two--channel Kondo model behaves in an 
analogous fashion. 

Within the NCA we here investigate case (ii) by  
examination of the behavior of the self-energies in equilibrium.
As mentioned above, within CFT\cite{afflud} the retarded 
conduction electron self--energy
$\Sigma_c(\omega,T)$ of the two--channel particle--hole symmetric
Kondo model is found to obey scaling of the form
\begin{equation}
\mbox{Im}\Sigma_c(\omega,T) - \mbox{Im}\Sigma_c(0,T)  =  b T^{1/2} 
H( \frac{\hbar\omega}{k_{B}T}) \,\, ,
\label{scalself}
\end{equation}
where $H$ is  again a universal scaling function
of the kind introduced above ($\eta=1/2$). 
The constant, $b$, is non-universal.  Within CFT the sign of $b$
the sign depends on whether one is on the
weak coupling or the strong coupling side of the (intermediate coupling)
fixed point\cite{afflud}. 
The NCA calculation is on the weak coupling side and yields
a positive constant $b$ (see below), in agreement with CFT.
However, a direct quantitative comparison with CFT is less
straight--forward because the model used in the CFT calculation is 
particle--hole (p--h) symmetric, while ours is not: The local level
$\epsilon _d$ has a  finite position 
\begin{figure*}[h]
\leavevmode\centering\psfig{file=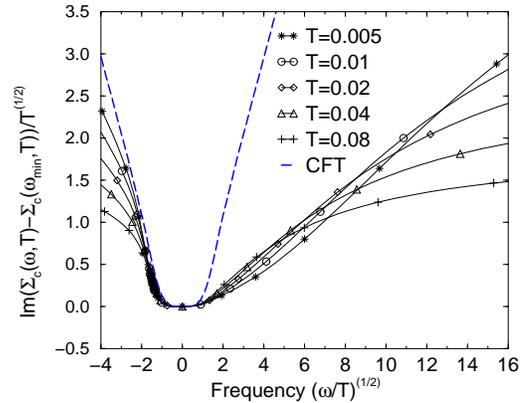,width=2.7 in}
\caption{Scaling plot for the imaginary part of the  retarded conduction  
electron self-energy for a small concentration ($x= 1\%$) of
the $M=2$ channel Anderson impurities in a noninteracting metal. 
Temperatures are given in units of $T_K$.
$\mbox{Im} \Sigma_c$ has a minimum that is shifted to positive frequencies
due to finite temperature effects. The data are scaled with respect to the
point ($\omega_{min},\mbox{Im}\Sigma(\omega_{min})$. For frequencies 
below $\omega_{min}$ the self-energy behaves like $(\omega/T)^{1/2}$
and scales well up to frequencies of the order of $T_K$. However,
for positive frequencies the self-energy is strongly temperature dependent and
scaling is less perfect. The parameters of the CFT prediction 
for the particle--hole symmetric Kondo 
model (dashed line) have been adjusted so that the slope for negative 
arguments matches that of the lowest temperature NCA curve.}
\label{sigccft}
\end{figure*}
\noindent
below the Fermi level, while
the on--site repulsion is taken to be infinite. 
The  scaling plot Fig. \ref{sigccft} 
shows a very  strong asymmetry about the point $\omega =0$.
The temperature dependence is also substantial. 
The parameters of the CFT curve (dashed line) have been adjusted 
so that the slope for
the linear part in $(\omega/T)^{1/2}$ for negative frequencies matches the 
slope of the lowest $T$ NCA curve. 
The asymmetry of the conduction electron self--energy may be traced
back to the p--h asymmetry of our model. P--h asymmetry
may very well be present in the experimental systems\cite{rlvdb}.
However, no or only small asymmetries are seen in the measurements of 
the nonlinear conductance. This is presumably because in the 
expression for the nonlinear current, Eq.~(\ref{curr}), the frequency 
is integrated over both positive and negative values, $-V/2 \leq
\omega \leq +V/2$, thus averaging out the asymmetry. This conjecture 
is supported by our calculation of the conduction electron
self--energy (Fig.~\ref{sigccft}), which displays strong p--h
symmetry, and the non--linear conductance\cite{hkh1}
(Fig.~\ref{scal12}), which is almost p--h symmetric.

Finally, in the Anderson model 
we also consider the impurity electron self-energy 
$\Sigma_d (\omega,T)$.  It can easily be computed from the 
d--Green function.
The two physical self-energies $\Sigma_d (\omega,T)$ and
$\Sigma_c (\omega,T)$ are nonlinearly related via Eqs.~(\ref{sigceq}),
(\ref{tmatrix}) for a system of dilute impurities in an equilibrium situation. 
It should be noted that the nonlinear conductance is directly related to
the spectral function $A_d(\omega)$ via Eq. \ref{curr}. We
first examine whether the impurity self-energy $\Sigma_d (\omega,T)$
shows scaling behavior {\it in equilibrium} in the variables $\omega$ and $T$.
The imaginary  
\begin{figure*}[t]
\leavevmode\centering\psfig{file=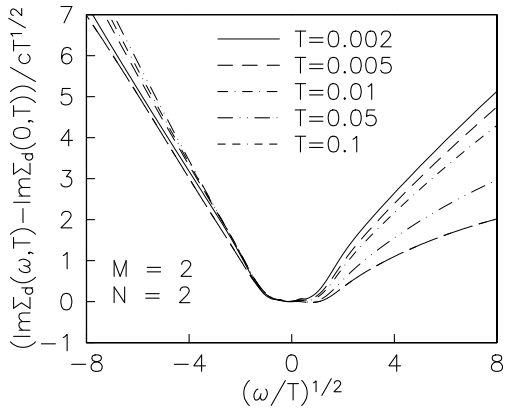,width=2.7 in}
\caption{Scaling plot for the imaginary part of the impurity electron 
self-energy for the $M=2$ channel Anderson model. Temperatures are given in
units of $T_K$.
For different temperatures $T$, 
$\mbox{Im}(\Sigma_d (\omega,T) - \mbox{Im}\Sigma_d (0,T))/(c T^{1/2})$ is 
plotted versus the square root of the scaled frequency, $(\omega/T)^{1/2}$. 
The constant $c$ depends on details of the model.
The left parts of the curves
($\omega < 0$) obey the anticipated square root behavior
and scale very well for $|\omega| \ll T_K$.
For $\omega/T > 0$ the NCA curves show
a strong $T$--dependence even for $T \ll T_K$. This is a possible 
origin of the 
$T$ dependent slopes of the nonlinear conductance curves in Fig. 5.
However, for modestly large
frequencies, e.g. $\omega/T < 4$ the lowest $T$ curves seem to follow 
square root behavior, too. The general asymmetry of the self-energy 
is a consequence of the particle--hole asymmetry of the Anderson
model considered here.}
\label{sigd}
\end{figure*}
\noindent part of the 
retarded impurity self-energy is negative by causality
and its absolute value shows a peak at the Fermi level.
In Fig. \ref{sigd}, we plot
$(Im \Sigma_d (\omega,T) - Im \Sigma_d (0,T))/(c T^{1/2})$ vs. 
$(\omega/T)^{1/2}$  for different temperatures $T$.
The constant $c$ is positive and depends on the parameters of the model.
Obviously, the left parts of the curves
($\omega < 0$) obey scaling very well. For arguments of 
$(\omega/T)^{1/2}$ outside the scaling regime $\omega, T < T_K$
the curves bend, and the self-energies
grow roughly logarithmically (not shown in the figure). 
In contrast, the $\omega/T > 0$ parts of the NCA--curves show
a strong $T$--dependence even for $T \ll T_K$, again a consequence of 
p--h asymmetry. 
Nevertheless, at negative frequencies scaling with the 
correct power law is established over a wide range of the scaling argument
$(\omega/T)^{1/2}$, as long as $\omega, T < T_K$, even though $\omega$ can be
much larger than $T$. It is very plausible 
that this scaling is reflected in the  scaling of the nonlinear conductance
in  the arguments bias and temperature as observed in experiment\cite{rlvdb}
and our numerical evaluation.\cite{hkh1} Furthermore, the deviations from 
scaling at positive frequencies could be another reason for the observed
finite $T$--corrections to scaling of the conductance.\cite{hkh1}

\subsection{Conductance with asymmetric couplings}
Up to this point we have taken the couplings of the impurity to the
conduction bands to be equal, $\Gamma_L = \Gamma_R$. As mentioned before,
especially for a tunnel junction there is no reason why this should be the
case. 
The NCA--Eqs. (\ref{ncaret}), (\ref{ncales}) are \it not \rm
symmetric in the couplings, that is $\Gamma_L \leftrightarrow \Gamma_R$ is not
a symmetry of the equations. This suggests that the differential
conductance signals
are not symmetric about zero bias if $\Gamma_L \neq \Gamma_R$.
Indeed,
the Onsager relations for a two terminal measurement 
only apply to the linear response regime.  For nonlinear
response there is no simple relation between $I(V)$ and $I(-V)$.
However, interchanging both
$\Gamma_L \leftrightarrow \Gamma_R$ {\it and}
$V \leftrightarrow -V$ is a symmetry. It is therefore enough to show only the
conductances for $\Gamma_L > 1/2$. The curves with $\Gamma_L < 1/2$
can be obtained from the $\Gamma_L > 1/2$ ones by reflection about the y-axis.

\begin{figure*}[b]
\leavevmode\centering\psfig{file=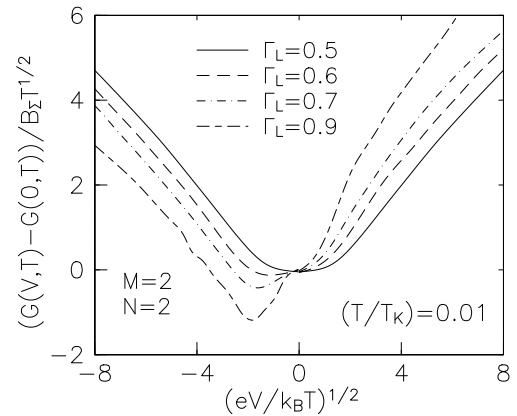,width=2.7 in}
\caption{Nonlinear conductance for the $M=2$ channel case for asymmetric 
coupling, $\Gamma _L \neq \Gamma _R = 1 - \Gamma _L$. 
As expected from the asymmetry of the NCA equations (16) and (17),
the conductance signals show a quite strong asymmetry about zero
bias even for moderate differences in the couplings. Asymmetries in  the 
conductance have been observed in metal--insulator--metal tunnel junctions.}
\label{asymm}
\end{figure*}

An example of such asymmetric conductance curves is shown in Fig. \ref{asymm}.
The data is for the two--channel model, but the qualitative
aspects of asymmetry does not 
depend on the channel number. The constant $B_{\Sigma}$ is dependent on the
asymmetry, but has been divided out for better comparison of the curves.
The asymmetry is pronounced even for moderate deviations from symmetric 
coupling. 
Asymmetric conductance vs. voltage curves similar to those shown 
in Fig. \ref{asymm} have been observed in Ta-I-Al
tunnel junctions,\cite{app-shen}
where they were plotted as an odd in voltage contribution to the
differential conductance.
\subsection{Dynamic and static susceptibility for the two channel model}
Finally, we also show results for the static and dynamic susceptibilities 
with and without finite bias. 
Although it is unlikely that one will be able to measure the
susceptibility for a single impurity, the susceptibility is one of 
the clearest measures of the screening of the impurity by electrons.
All data shown below is for the two channel 
\begin{figure*}[h]
\leavevmode\centering\psfig{file=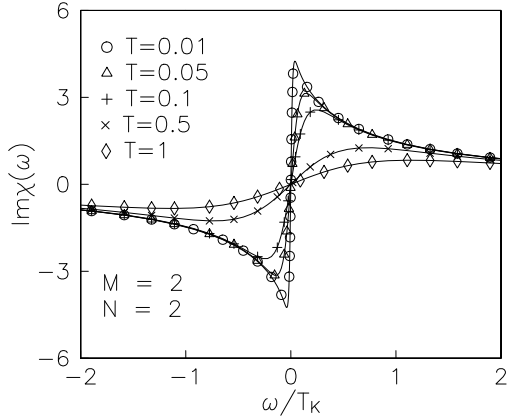,width=2.7 in}
\caption{Imaginary part of the dynamic susceptibility ${\rm Im}\chi (\omega)$
(arb.~units) for the $M=2$ channel Anderson model in equilibrium 
($V=0$). Temperatures are given in units of $T_K$. For $T\rightarrow 0$, 
$\chi(\omega)$ 
behaves as 
$\mbox{Im}\chi(\omega) \simeq c_1 \mbox{sign}(\omega)
[1-c_2\sqrt{|\omega/T_K|}]$,
in agreement with exact results for the two--channel case. This is in
contrast to the exact linear behavior in the $M=1$ channel model. }
\label{imchom}
\end{figure*}
\noindent model. The 
results for the usual Kondo model show different power laws, but the
general effect of the finite bias is the same.\\

In equilibrium,  in the zero temperature limit, the dynamic susceptibility 
defined in Eq. (\ref{suscep}) is given by
a step function of the form \cite{cox3}
\begin{eqnarray}
\mbox{Im}\chi(\omega) = c_1 \mbox{sign}(\omega)\biggl[ 1-c_2\sqrt{\bigl| 
\frac{\omega}{T_K}\bigr|}
+ \mbox{\ldots}\biggr]\,\,\, .
\label{ncasus}
\end{eqnarray}
The NCA approaches this behavior as the temperature is reduced.
At finite temperature, the step is  broadened, as shown in
Fig.~\ref{imchom}, with
the extrema located at values 
\begin{figure*}[h]
\leavevmode\centering\psfig{file=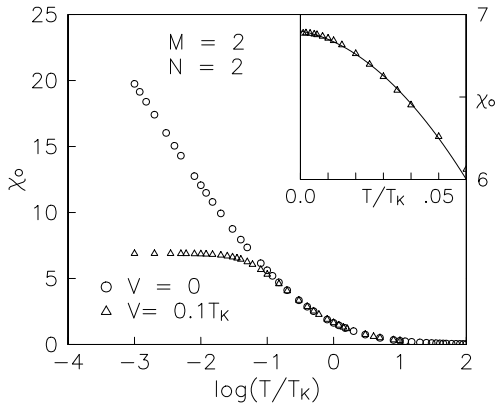,width=2.7 in}
\caption{Static susceptibility $\chi_o$ (arb.~units) vs. temperature at zero and 
at finite bias $V$ for $M=2$, $N=2$.
In equilibrium, $\chi_o$ shows the characteristic, expected
logarithmic divergence as $T$ approaches zero for the two--channel model.
Out of equilibrium, this
divergence is cut off at a temperature corresponding to the bias $V$. 
The inset shows that $\chi_o$ falls with $T^2$ below this cutoff.
For high temperatures $T \gg T_K$, $\chi_o$ falls like $1/T$ 
(Curie--Weiss law).}
\label{rechvst}
\end{figure*}
\begin{figure*}[b]
\leavevmode\centering\psfig{file=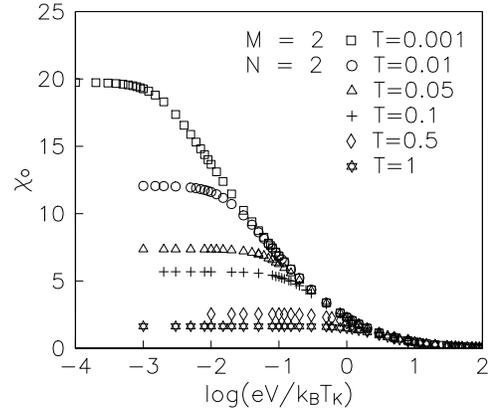,width=2.7 in}
\caption{Static susceptibility $\chi_o$ (arb.~units) vs. bias $V$
at various temperatures $T$ for $M=2$, $N=2$. Temperatures are given in 
units of $T_K$.
$\chi_o$ has a very similar dependence on $V$ and $T$ as long as $V,T < T_K$
(scaling regime).
$\chi_o$ drops like $V^2$ for $T\buildrel <\over\sim0.1 T_K$
and like log$(V)$ around $T_K$.
However, for large  $V \gg T_K$, $\chi_o$ falls less rapidly with $V$ 
than with $T$, see Fig.~11.}
\label{rechvsv}
\end{figure*}
\noindent which grow roughly with $T^{1/2}$.
The real part follows by a Kramers-Kroenig relation and diverges
logarithmically for $\omega\rightarrow 0$, again cut off at finite $T$. 
As a consequence, the static susceptibility $\chi_o = {\rm Re}\chi (\omega =0)$ 
diverges logarithmically as $T$ approaches zero in agreement 
with non--Fermi liquid behavior, as has been predicted before.
\cite{bethemulti,afflud,cox3}  This logarithmic divergence is 
well reproduced by the NCA technique, see Fig. \ref{rechvst}\\ 

Out of equilibrium, the finite bias serves as another low energy cutoff, but in
a nontrivial manner. If we look at the extrema of the imaginary part of the 
susceptibility 
\begin{figure*}[h]
\leavevmode\centering\psfig{file=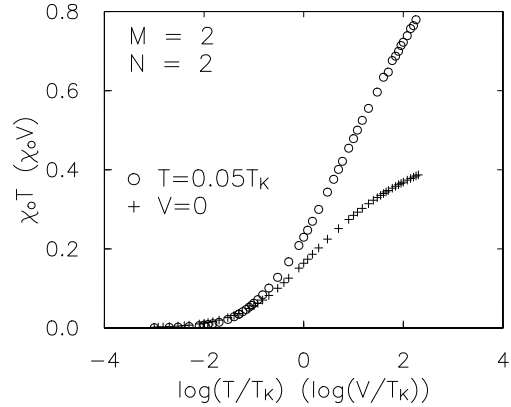,width=2.7 in}
\caption{Product of the static susceptibility $\chi_o$  and temperature $T$ 
(bias $V$) vs. $T$ ($V$) on a semi--logarithmic scale for $M=2$, $N=2$.
The $T$ dependence shows saturation at high temperatures
and therefore implies the Curie law, $\chi_o \propto 1/T$. 
However, the $V$ dependence is linear at large bias, implying
that $\chi_o$ falls less rapidly with $V$ 
than with $T$, $\chi_o \propto \mbox{log} (V)/V$.  
The y--axis units are such that a ``free pseudo--spin'' would correspond 
to a constant value of 1/2 (Curie--behavior at large temperatures).}
\label{chitv}
\end{figure*}
\noindent 
at low temperature but finite bias ($V >T$) we find that 
they are located at smaller absolute values than at the 
corresponding temperature. The logarithmic 
divergence of the Re$\chi (\omega)$ is cut off at 
about $V$, so that the static 
susceptibility does not diverge logarithmically as $T\rightarrow 0$ anymore. 
Instead, it approaches a ($V$--dependent) finite value with a quadratic 
$T$--dependence (see inset of Fig. \ref{rechvst}).
However, this does not signal the return of Fermi liquid behavior for $T < V$,
since we still have $\sqrt{V/T}$ behavior of the conductance for $V$ well
below $T_K$. Fig. \ref{rechvst} shows  the $T$--dependence of $\chi_o$
for $V= 1/10 T_K$. 
\begin{figure*}[h]
\leavevmode\centering\psfig{file=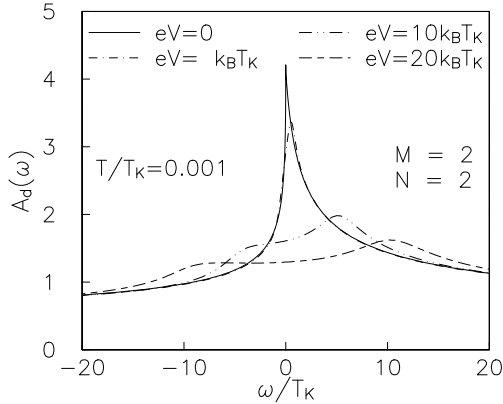,width=2.7 in}
\caption{The impurity spectral function $A_{d}(\omega)$ 
for the $M=2$ channel, $N=2$ Anderson model for various
biases $V$. The half width at half maximum of the zero bias spectral function 
(solid line) is a measure of $T_K$.
As the voltage is increased to $eV=k_{B}T_{K}$ the Kondo resonance
is reduced (dash-dotted line).  At very large bias the resonance
shows a shoulder and eventually splits into two peaks. Increasing the
temperature would wash out the peak splitting and restore a single,
much less pronounced peak.}
\label{adfig}
\end{figure*}

Similar behavior (i.e. quadratic in $V$ for low $V$, logarithmic for 
$T < V < T_K$) is observed for the $V$--dependence of 
the static susceptibility (see Fig. \ref{rechvsv}); 
however, there is a difference in the dependence on $T$ and $V$ in the 
regime $T_K < T,V$ and $T,V < \Gamma_{tot}$. For large $T>T_K$ at zero bias 
the static susceptibility 
behaves like $1/T$, indicating Curie-Weiss behavior. However, for large $V$
at low temperature
$\chi_o$ falls less rapidly.
The difference becomes  obvious if we plot $V\chi_o$ vs. $\log (V)$ 
and $T\chi_o$ vs. $\log (T)$
as shown in Fig. \ref{chitv}. Whereas the $T$-dependence saturates, 
indicating the free moment at high temperatures the $V$-dependence 
shows linear behavior, leading to $\chi_o \sim b\log(V)/V$. 
This stresses again
the different consequences of rising $T$ and $V$ once one has left the
scaling regime $T,V < T_K$.
The latter becomes most obvious 
if we look at the impurity spectral function $A_d(\omega)$
for various values of the bias, see Fig. \ref{adfig}. For zero bias and low 
temperatures there is a sharp resonance with width $T_K$ (solid line).
Increasing the temperature above $T_K$  the peak would broaden at the expense 
of its height (not shown). In contrast, if we keep the temperature low,
$T \ll T_K$, and increase the bias, the resonance first develops a shoulder
and then splits into two much broader but distinguished peaks
(see $eV = 10k_BT$ and $20k_BT$).
Increasing the
temperature would eventually wash out the peak splitting and restore a single,
though much less pronounced peak. This difference in behavior at large $T$
vs. large $V$ is the reason for the breakdown of scaling of the conductance
for $T$ or $V$ larger than $T_K$
and the different behavior of the susceptibility at large arguments
(Fig. \ref{chitv}).
\section{Conclusion}
In conclusion,
we have 
described in detail the analytical foundations and the numerical implementation
of the NCA integral equations for the one-- and two--channel Anderson model
out of equilibrium. Our algorithms enabled us to  reach lower temperatures
than previously obtained, allowing us to study the physics deep inside the
scaling regime of the two--channel model.

\par In linear response, we computed  the conductance for
tunnel junctions and point contacts
as well as the bulk resistivity. The two--channel data
for both properties show $T^{1/2}$ behavior in agreement with results
obtained by other methods. For the single--channel model  and $N=2$ we find 
dominantly linear behavior below $T_K$. 
For $N=6$ the bulk resistivity
drops with $T^2$ (Fermi liquid behavior) at the lowest temperatures considered 
in this work; however, the tunnel junction
conductance \sl rises \rm with $T^2$, reaches a maximum below the Kondo
temperature $T_K$ and then falls off logarithmically at higher $T$. 
This ``hump''
is associated with the fact that the Kondo peak of the impurity spectral
function is shifted away from the Fermi level for values of $N > 2$.

If we turn on a finite bias $V$, the Kondo peak of the impurity 
spectral function first diminishes in height and broadens, then splits
into two peaks located at the energies of the two Fermi levels of the leads
at a bias of about $10 T_K$.
The non--equilibrium conductance is again consistent with linear behavior 
in the regime $T< V < T_K$ for the single--channel
case with $M=1$, $N=2$. Therefore, we can plot the  conductance as a function of 
$eV/k_BT$ and achieve scaling for modest bias $V$.  Whether similar scaling 
of the conductance but with argument $(eV/k_BT)^2$ can exist for the case
$N=6$ is yet to be determined. The tunnel junction conductance falls with 
$V^2$  for bias $V< T_K$. 
This is in stark contrast to the hump in the
$T$--dependence of the zero bias conductance. If at all, scaling seems
possible only for temperatures well below the temperature where the hump 
occurs. 
The two channel data show scaling with respect to the argument  $(eV/k_BT)^{1/2}$
consistent with conductance measurements on clean point contacts. It has to be
pointed out, though, that the scaling at non--zero bias in the two--channel as well 
as in the single--channel
model is only approximate. Finite $T$ corrections are observed in the
numerical data (and also in the experimental data) for temperatures down
to about $1/100\, T_K$. This indicates that finite bias $V$ and finite temperature
$T$ are not equivalent, although they have qualitatively similar effects on
the conductance.
The scaling of the conduction electron self-energy turns out to be worse than 
that of the nonlinear conductance. This may be traced back to the lack 
of particle--hole symmetry of our model, which leads to asymmetries in 
the self-energy even at the lowest temperature. Additionally, there
are  also strong temperature dependent corrections to the square root behavior.

If we allow for asymmetric couplings  to the left and right Fermi seas,
we observe  conductance signals which are asymmetric about 
zero bias. Such features have been  seen in experiments on
metal-insulator-metal tunnel junctions. 

Finally, we
also calculated the dynamic and static (pseudo--) spin susceptibility 
and discussed the 
modifications due to a finite bias by example of the two channel model.
The dynamic susceptibility approaches a finite step at $\omega =0$ 
as $T\rightarrow 0$, leading to a logarithmic divergence of the static 
susceptibility in agreement with CFT results.
A finite bias cuts off this logarithmic divergence. In a very similar
fashion the temperature cuts off the divergence as the bias is vanishing.
Differences in the bias and temperature dependence of the static 
susceptibility appear at high bias and temperature outside of the 
scaling regime.\\

It is a pleasure to acknowledge discussions with
V.~Ambegaokar, R.~Buhrman, J.~von Delft, D.~Ralph, S.~Upadhyay, L.~Borkowski 
P.~Hirschfeld, K.~Ingersent, A.~Ludwig, A.~Schiller, and M. Jarrell.
This work was supported by NSF grant DMR-9357474, the U.F.~D.S.R., 
the NHMFL (M.H.H., S.H.), by NSF grant DMR-9407245, the German DFG 
through SFB 195 and by a
Feodor Lynen Fellowship of the Alexander v. Humboldt Foundation (J.K).
Part of the work of J.K. was performed at LASSP, Cornell University.

\appendix
\section{Numerical implementation of the NCA equations}
Below we briefly review the slave boson projection
technique and describe an implementation which allows for a highly
accurate as well as efficient numerical treatment of the singularities
of the spectral functions, which arise from the projection.

The exact projection of the expectation value of any operator $\hat{\cal O}$
onto the physical subspace, $Q=1$, 
is achieved by first taking the statistical average in the grand
canonical (GC) ensemble with a chemical potential $-\lambda$ for both
fermions $f$ and bosons $b$, and then differentiating w.r.t. the
fugacity $\zeta = {\rm exp}(-\beta\lambda )$ and taking the limit 
$\lambda \rightarrow \infty$,
\begin{eqnarray}
\langle \hat{\cal O}\rangle _C &=&\lim _{\lambda\rightarrow\infty}
\frac{\frac{{\rm d}}{{\rm d}\zeta} tr [\hat{\cal O} {\rm e}^{-\beta
      (H+\lambda Q)}]}
     {\frac{{\rm d}}{{\rm d}\zeta} tr [{\rm e}^{-\beta
      (H+\lambda Q)}]}\\
&=& 
\frac{\lim _{\lambda\rightarrow\infty}\langle\hat{\cal O}\rangle _{GC}
      {\rm e}^{\beta\lambda}}
     {\lim _{\lambda\rightarrow\infty}\langle Q\rangle _{GC}
      {\rm e}^{\beta\lambda}}
\end{eqnarray}
Note that in this expression the factor $Q$ arising from the
differentiation ${\rm d}/{\rm d}\zeta$ in the numerator may be dropped
for any operator ${\cal O}$ whose expectation value in the subspace
$Q=0$ vanishes (like, e.g., ${\cal O}= d_{\sigma\tau}(t) 
d^{\dag}_{\sigma\tau}(t')$ or any other {\it physically observable}
operator on the impurity site). The canonical (C) partition function
is given by 
\begin{eqnarray}
Z&=&\lim _{\lambda\rightarrow\infty} \left[ {\rm e}^{\beta\lambda}
\langle Q\rangle _{GC}(\lambda )\right]\, Z_{Q=0}\\
&=&Z_{Q=0}\, \int d\epsilon {\rm e}^{-\beta\epsilon}
(N A(\epsilon )+M B(\epsilon )) \, ,
\label{Zcanon}
\end{eqnarray}
where $Z_{Q=0}$ is the partition function in the subspace $Q=0$.

The integrals involved in the NCA equations 
are difficult to compute because of the singular
threshold structure\cite{muhart} of $A(\omega)$, $B(\omega )$, where
the position of the threshold energy $E_o$ is {\it a priori} not
known. In order to make the numerical evaluations tractable, we
apply a time t dependent 
$U(1)$ gauge transformation simultaneously to the $f$ and $b$
particles according to $f\rightarrow {\rm exp}(i\lambda _o t) f$,
$b\rightarrow {\rm exp}(i\lambda _o t) b$. This transformation is
a symmetry of the Anderson model and amounts to a shift of the
slave particle energy or chemical potential by $\lambda _o$,
$\omega \rightarrow \omega +\lambda _o$.
Note that this shift does not affect any physical properties, as seen
explicitly, e.g., from Eq.~(\ref{specfun}). After this energy shift,
the spectral and lesser functions appearing in the NCA equations
read
\begin{mathletters}
\begin{eqnarray}
A_{\lambda_o}(\omega) &=& \frac{\mbox{Im} \Sigma^r(\omega)}{(\omega - \epsilon_d
 + \lambda_o - \mbox{Re}\Sigma^r(\omega))^2 + (\mbox{Im}\Sigma^r(\omega))^2}\\
B_{\lambda_o}(\omega) &=& \frac{\mbox{Im} \Pi^r(\omega)}{(\omega  +
\lambda_o - \mbox{Re}\Pi^r(\omega))^2 + (\mbox{Im}\Pi^r(\omega))^2}
\end{eqnarray}
\label{aspeclam}
\end{mathletters}
\begin{mathletters}
\begin{eqnarray}
a_{\lambda_o}(\omega) &=& \frac{\Sigma^<(\omega)}{(\omega - \epsilon_d +
\lambda_o - \mbox{Re}\Sigma^r(\omega))^2 + (\mbox{Im}\Sigma^r(\omega))^2}\\
b_{\lambda_o}(\omega) &=& \frac{\Pi^<(\omega)}{(\omega  +
\lambda_o - \mbox{Re}\Pi^r(\omega))^2 +
(\mbox{Im}\Pi^r(\omega))^2}\,\,\, .
\end{eqnarray}
\label{alesslam}
\end{mathletters}

\noindent
In particular, we now have from Eq.~(\ref{Zcanon})
\begin{eqnarray}
\frac{Z(\lambda _o)}{Z_{Q=0}}={\rm e}^{-\beta\lambda_o} 
\int d\epsilon {\rm e}^{-\beta\epsilon}
(N A(\epsilon )+M B(\epsilon )) \, ,
\label{Zshifted}
\end{eqnarray}
The crucial point about making the numerics efficient is that
$\lambda_o$ is determined {\it in each iteration} such that the
integral in Eq.~(\ref{Zshifted}) is equal to unity\cite{kroha,costi}. 
This definition of $\lambda _o$
forces the zero of the auxiliary particle energy to coincide with 
the threshold energy, $E_o=0$ in each iteration step. Thus, it 
enables us to define fixed frequency meshes, which do not change 
from iteration to iteration and at the same time resolve the 
singular behavior very well, as described in appendix B.  
The procedure described above leads to a substantial gain in precision
and significantly improves the convergence of the iterations,
even though the equation determining $\lambda _o$ must be solved 
during each iteration. 

From Eq.~(\ref{Zshifted}) and the definition of the impurity
contribution to the free energy, $F_{imp}(T)$, 
${\rm exp}(-\beta F_{imp})=Z(T)/Z_{Q=0}(T)$, it is seen that $\lambda
_o$ determined in the above way is just equal to $F_{imp}$. This
provides a convenient way of calculating  $F_{imp}(T)$ directly from
the auxiliary particle Green functions.

\section{Integration meshes for equilibrium and non-equilibrium NCA}
The various features of the auxiliary particle as well as the physical 
spectral functions are
charcterized by energy scales, which differ by several orders of
magnitude. These energy scales are the conduction band
width $D$, the localized level $\epsilon _d$, and the dynamically
generated Kondo scale, $T_K$, which is typically of order $10^{-4}D$. 
Moreover, because of the 
$T=0$, $V=0$ threshold divergence of the auxiliary particle spectral
functions, the sharpest features have a width given by the 
temperature, which can be of the order of $10^{-7}D$. In
non--equilibrium, the bias $V$ appears as an additional scale.
In the numerical solution of the NCA equations, discrete, non--equidistant
intregration meshes must be set up such that all the features at the
various energy scales are well resolved.  
  
These meshes can be generated by mapping the grid points $x_i$ of an
equidistant mesh onto the non--equidistant 
frequency points $\omega _i$ by means of an appropriately chosen 
function $h(x)$. In the regions where the very    
sharp features of the spectral functions and the Fermi function
appear, i.e.~near 
$\omega =0$ and $\omega = \pm V/2$, respectively, we will use
a logarithmicly dense mesh. On the other hand, 
in order to resolve the relatively broad
peak centered around the local level $\varepsilon _d$, the
substitution $\omega _i = \varepsilon_d +c \tan (x_i)$ will be used.

In general, the entire interval of integration is composed of
$L$ meshes $\{ x^l_i \}$, $i=1\dots n_l$, $l=1\dots L$. We map these
meshes onto the nonuniform frequency meshes $\{ \omega ^l_i \}$ via
\begin{eqnarray}
\omega_i^l = h^l (x^l_i) , \,\,\,\,\,\,\,\, i=1\ldots n^l.
\end{eqnarray}
We can now 
rewrite the integration of an arbitrary function $k(\omega)$ as
an integration over the ``equidistant'' variables $\{x_i^l\}$:
\begin{eqnarray}
\int_{-\infty}^{\infty} &d\omega& k(\omega)= 
\sum_l \int_{a_l}^{b_l} dx \frac{\partial h^l (x)}{\partial x}
k(h(x)) \nonumber  \\ 
&\simeq & \sum_l {\Delta x^l} \left(  \sum_{i=2}^{n_l -1}  \left[
\frac{\partial h^l}{\partial x^l}(x^l_{i}) k(h(x^l_{i}))\right]\right.\\
&+& \left.\frac{1}{2}
\left[\frac{\partial h^l}{\partial x^l}(x^l_{1}) k(h(x^l_{1})) +
      \frac{\partial h^l}{\partial x^l}(x^l_{n_l}) k(h(x^l_{n_l}))
\right]\right).\nonumber
\end{eqnarray}
The $a^l =\omega^l_1, b^l = \omega^l_{n_l}$ are the limits of integration of 
the different regions of the frequency--axis. 
To cover the whole axis we must have $a^{l+1}=b^l$.
In equilibrium, we can get by with four regions: $[-\infty , -\omega_I),\,
[-\omega_I,0=\epsilon_F),\,[0,\omega_I),\,[\omega_I,\infty]$ , where
$\omega_I$ is an  interface frequency where two regions of the mesh
are matched. ($|\epsilon_d| - \Gamma > \omega_I >> T_K$). 
By choosing the functions 
$h^l(x^l)$ as  $\varepsilon_d+c_1\tan (x^l)$ 
in the regions with large absolute frequency
and as  $c_2 \exp (x^l)$ in the regions $|\omega| < \omega_I$ we create
large mesh point spacings far from 
$\epsilon_F$ and exponentially small spacings (``logarithmic'' mesh)
at  $\epsilon_F = 0$. Proper 
adjustment of constants in the $h^l$'s is required. 
The frequency mesh point spacing near $\omega = 0$ 
should be at least 10 times 
smaller than $T$ (and/or $V$ out of equilibrium). 
Crucial for the success of this procedure is the introduction 
of $\lambda_o$ (see Appendix A) in the iteration procedure. $\lambda_o$
shifts the peaks of the slave particle functions to the neighborhood 
of $\omega =0$ in each iteration step. This allows to define
a fixed frequency grid, which leads to a significant increase in
computational speed and precision. 

Out of equilibrium the distribution function is a double step function
with steps at $\pm V/2$. It turns out that in the Kondo limit the
slave boson spectral and lesser functions show broadened peaks
at about the same frequencies. However, the pseudo--fermion functions 
behave differently. They do {\it not} split, 
but have a single peak somewhere between the Fermi level and 
$V/2$ that shifts not linearly with $V$.
To cope with such behavior we wish to have good resolution
at $\pm V/2$ and at $\epsilon_F$. (The latter one is to 
improve the resolution at the location of the peak of the
pseudo-fermion functions. Unfortunately, we do not know how
this location will move with increasing $V$) . 
To achieve this we let the logarithmic
mesh end at $\pm V/2$ coming from larger/smaller frequencies and choose
the spacing in between according to the sum of two tanh--functions
which have their zero shifted to $\pm V/4$, respectively. We have to
choose parameters of these functions, so that the mesh spacings at the crucial
energies is small enough to resolve all features of the integrand.
These parameters depend on the bias $V$. They have to be calculated before the
mesh is 'set up' whenever we change the potential from one run to the next.
However, once the mesh is set, we do not have to change it anymore during
the iterations, because of the same reasons as in equilibrium.

The typical total 
number of integration points used is 200 and 250 for equilibrium
and out of equilibrium, respectively. Out of equilibrium we need about 50
points more for the `inner' region between $\pm V/2$ at moderate bias 
$V < 20 T_K$. For higher bias we have to introduce more points in the inner
region. Convergence is achieved within 100 - 200 iterations. The 
CPU time to obtain a converged solution on a typical workstation
is below 1 minute for the equilibrium case, and of the order of
minutes for non--equilibrium.

\section{General formula for the conductance}
In this appendix we derive Eq. (\ref{curr}) for the current through a
constriction with an impurity.  We proceed in three stages.
First, we introduce our scattering state notation and 
review the noninteracting case.  Next, we derive a general
formula for scattering from an interacting impurity.
This is valid for point contacts, tunnel junctions, and
anything in between.  Finally, we specialize to the case
of a clean point contact. 

The geometry we consider consists of perfect left ($L$)
and right ($R$) leads connected by a central region
where there is scattering.  The scattering states,
$\psi ({\bf x})$, are eigenstates of the noninteracting 
problem.  They are labeled by their incoming wave vectors,
${\bf k}$, where $k_z>0$ corresponds to a right moving wave
and $k_z<0$ corresponds to a left moving wave,
where $z$ is the direction along the length of the leads.
For example, a state moving from left to right ($k_z>0$)
has the asymptotic form for $z\gg 0$ of
\begin{equation}
      \psi _{\bf k}({\bf x})
	 =\sum _{{\bf k}_{\perp}'} t_{{\bf k}'{\bf k}}^{RL}\,
	       \sqrt{|v_z/v_z'|} \,e^{i|k'_z|z}\,
 	       \varphi _{{\bf k}_{\perp}'} ({\bf x}_{\perp})
          \label{asymr}
\end{equation}
and for $z\ll 0$ of
\begin{eqnarray}
      \psi _{\bf k}({\bf x})
	&=&e^{i|k_z|z}\,\varphi _{{\bf k}_{\perp}} ({\bf x}_{\perp}) 
              \label{asymr2} \\
	      &+&\sum _{{\bf k}_{\perp}'} r_{{\bf k}'{\bf k}}^{L}\,
	       \sqrt{|v_z/v_z'|} \,e^{-i|k'_z|z}\,
 	       \varphi _{{\bf k}_{\perp}'} ({\bf x}_{\perp})
              \nonumber
\end{eqnarray}
The transverse modes,
$\varphi _{{\bf k}_{\perp}} ({\bf x}_{\perp})$, in Eqs. (\ref{asymr})
and (\ref{asymr2}) are chosen to have unit normalization,
and $v_z = k_z/m$ is the velocity along the length of the leads.
It is also understood that the energy of the incident and
transmitted waves are the same,
$\epsilon _{{\bf k}_\perp} + \epsilon _{k_z}
 = \epsilon _{{\bf k}'_\perp} + \epsilon _{k'_z}$.

The current for both the interacting and the noninteracting
case may be expressed as a cross-sectional integral
of the `lesser' Green function:
   \begin{equation}
      I = \int \frac {d\omega}{2\pi} \int d^2x_{\perp} 
        \left(\frac {\nabla _z - \nabla _{z'}}{2mi}\right)
        g_<({\bf x},{\bf x}';\omega )\Big| _{{\bf x}={\bf x}'}.
        \label{curdef}
   \end{equation}
For the noninteracting case, this Green function may be written
in terms of the scattering states as
   \begin{equation}
     g_<^o (x,x';\omega ) = \int \frac {dk_z}{2\pi} \sum _{{\bf k}_\perp}
         2\pi \delta (\omega - E_{\bf k}) \psi _{\bf k}({\bf x})
	 \psi _{\bf k}^*({\bf x}') f_{\bf k}(\omega ), \label{nonintg} 
   \end{equation}
where $f_{\bf k}(\omega )$ is a Fermi function at chemical potential
$\mu _L$ for $k_z>0$ and at $\mu _R$ for $k_z < 0$.  We will usually
refer to these Fermi functions as $f_L(\omega)$ and $f_R(\omega )$,
respectively.  Using the asymptotic expressions of 
Eqs. (\ref{asymr}) and (\ref{asymr2}) for the 
the right moving scattering
states and the similar ones for the left moving states, Eqs. 
(\ref{curdef}) and (\ref{nonintg}) lead to the usual Landauer formula
for the conductance:
   \begin{equation}
      I = \int \frac {dE_k}{2\pi} \sum_{{\bf k}_{\perp}{\bf k}'_{\perp}}
         |t^{RL}_{{\bf k}'{\bf k}}|^2 (f_L(E_k) - f_R(E_k)).
	\label{landauer}
   \end{equation}

We now add an impurity which includes an interacting term
 to the Hamiltonian.  The coupling of the
impurity, denoted by 0, to the electrons is given by 
   \begin{equation}
      H'= \int \frac {dk_z}{2\pi} \sum _{{\bf k}_\perp} 
  	 W_{\bf k} c_0^{\dag} c_{\bf k} + W_{\bf k}^*c_{\bf k}^{\dag}c_0,
         \label{coupling}
   \end{equation}
where ${\bf k}$ refers to the scattering state of incoming wave
vector ${\bf k}$, {\em not} a plane wave state.
In Eq. (\ref{coupling}) and in the previous equations we have not
included spin.  The entire derivation presented here follows through
in the presence of a spin (or other) index so long as the self-energy
is diagonal in that index.  This is the case for the Anderson model
used in this paper.  In order to simplify the notation, we shall 
proceed without spin and at the end quote the final result when the
electron spin is included.

Using Dyson's equation one can express all of the Green functions for the
full system, $g$, in terms of the noninteracting Green functions, $g^o$,
and the full Green function at the impurity, $g(0,0)$.
In particular the Green function
$g_<({\bf k},{\bf k'})$, which is used to compute the
current, is given by
   \begin{eqnarray}
      g_<({\bf k},{\bf k}') &=& 2\pi \delta (k_z - k_z')
                                   \delta _{{\bf k}_{\perp}{\bf k}_{\perp}'} g_<^o({\bf k}) \nonumber \\
         &+& g_<^o({\bf k})W_{\bf k}^*g_a(0,0)W_{{\bf k}'}g^o_a({\bf k}') \nonumber \\
         &+& g_r^o({\bf k})W_{\bf k}^*g_<(0,0)W_{{\bf k}'}g^o_a({\bf k}') \label{gl1} \\
         &+& g_r^o({\bf k})W_{\bf k}^*g_r(0,0)W_{{\bf k}'}g^o_<({\bf k}') \nonumber .
   \end{eqnarray}
In Eq. (\ref{gl1}) all Green functions have the same energy, $\omega$.
The self-energy $\sigma$
contains the many body interaction at the impurity site.
Equation (\ref{gl1}) is converted to real space using
   \begin{equation}
      g_r({\bf x},0) = \int \frac {dk_z}{2\pi} \sum _{{\bf k}_{\perp}}
		       \psi _{\bf k}({\bf x}) g_r({\bf k},0),
   \end{equation}
and the similar relation for the advanced Green function.
The result for the real space $g_<$ is
   \begin{eqnarray}
      g_<({\bf x},{\bf x}')
			   &=&  g_r({\bf x},0)\, \sigma _<\, g_a(0,{\bf x}') \label{grealspace} \\
                           &+&\int \frac {dk_z''}{2\pi}
                              \sum _{{\bf k}_{\perp}''}
			      \left\{ 
			         \psi _{{\bf k}''}({\bf x}) 
				 + g_r({\bf x},0) W_{{\bf k}''}
			      \right\} \nonumber \\ 
                           &\times &
			      g_<^o({\bf k}'')\,
			      \left\{ 
			         \psi _{{\bf k}''}^*({\bf x'}) 
				 + W_{{\bf k}''}^* g_a(0,{\bf x}') 
			      \right\} . \nonumber 
   \end{eqnarray}
\noindent
As for the noninteracting case, we wish to evaluate the current
far into the left and right leads.  To do this we need the 
the asymptotic form of the scattering states 
(Eqs. (\ref{asymr}) and (\ref{asymr2})) and the asymptotic
form of the retarded and advanced Green functions, $g_{r(a)}({\bf x},0)$,
which we define as
   \begin{equation}
   \frac {g_r({\bf x},0)}{g_r(0,0)}
                     = \sum _{{\bf k}_{\perp}} t^{R(L)}_{\bf k}\, 
		          \varphi _{{\bf k}_\perp}({\bf x}_\perp )\,
		          e^{+(-)i|k_z| z}\Big| _{E_k=\omega} .
   \end{equation}
Substituting Eq. (\ref{grealspace}) into 
Eq. (\ref{curdef}) for the current, then yields
   \begin{eqnarray}
     I_R &=& \sum _{{\bf k}_{\perp},{\bf k}_{\perp}''}
            \int \frac {d\omega}{2\pi} 
	        |t^{RL}_{{\bf k}{\bf k}''} 
                 + t^R_{{\bf k}} g_r(0,0) W_{{\bf k}''}|^2
		f_L(\omega ) \label{rightcur}
	     \\
         &+& \sum _{{\bf k}_{\perp},{\bf k}_{\perp}''}
            \int \frac {d\omega}{2\pi}
	        (|r^{R}_{{\bf k}{\bf k}''} 
                 + t^R_{{\bf k}} g_r(0,0) W_{{\bf k}''}|^2
		 - \delta _{{\bf k}_{\perp}{\bf k}_{\perp}''})
		f_R(\omega ) \nonumber 
	      \\
         &+& \sum _{{\bf k}_{\perp}}\int \frac {d\omega}{2\pi} 
                v_z'' |t^R_{{\bf k}}g_r(0,0)|^2 \sigma _< (\omega ) \nonumber
   \end{eqnarray}
   \begin{eqnarray}
     I_L &=& -\sum _{{\bf k}_{\perp},{\bf k}_{\perp}''}
            \int \frac {d\omega }{2\pi}
	        |t^{LR}_{{\bf k}{\bf k}''} 
                 + t^L_{{\bf k}} g_r(0,0) W_{{\bf k}''}|^2
		f_R(\omega ) \label{leftcur} 
	     \\
         &-& \sum _{{\bf k}_{\perp},{\bf k}_{\perp}''}
            \int \frac {d\omega}{2\pi}
	        (|r^{L}_{{\bf k}{\bf k}''} 
                 + t^L_{{\bf k}} g_r(0,0) W_{{\bf k}''}|^2
		 - \delta _{{\bf k}_{\perp}{\bf k}_{\perp}''})
		f_L(\omega ) \nonumber 
	      \\
         &-& \sum _{{\bf k}_{\perp}}\int \frac {d\omega}{2\pi} 
                |v_z''| |t^L_{{\bf k}}g_r(0,0)|^2 \sigma _< (\omega ) \nonumber
   \end{eqnarray}
Eqs. (\ref{rightcur}) and (\ref{leftcur}) are our most general expressions
for the current.
It is useful to compare them to those 
for the noninteracting case (Eq. (\ref{landauer})).  Without the 
terms involving $\sigma _<$, equations (\ref{rightcur}) and 
(\ref{leftcur}) have exactly the same structure as the noninteracting
current.  The effect of the impurity is to change the transmission
probability for electrons coming from the left or the right.
The $\sigma _<$ contains the ``scattering-out" of an electron from the
impurity state.  This is a new feature of the interacting problem.

Equations (\ref{rightcur}) and (\ref{leftcur}) are valid for
an arbitrary scattering potential, including both the
tunnel junction case and the clean point contact case.
We model the clean point contact case by a perfect wire.
The wire will have a conductance equal to $e^2/h$ times
the number of channels at the Fermi energy.
The transmission and reflection probabilities for this
case are 1 and 0:
\begin{eqnarray}
\delta _{{\bf k _\perp},{\bf k _\perp ''}} &=&
   t^{RL}_{{\bf k },{\bf k ''}} =
   t^{LR}_{{\bf k },{\bf k ''}} \\
0 &=& 
   r^{R}_{{\bf k },{\bf k  ''}} =
   r^{L}_{{\bf k },{\bf k  ''}} .
\end{eqnarray}
In this perfect wire case the scattering states are
plane waves.
The impurity is placed at position, ${\bf x} = {\bf a}$,
and the overlap matrix elements are
\begin{equation}
W_{\bf k} = W^{L(R)} \frac {e^{ik\cdot a}}{\sqrt{{\cal A}}} ,
\label{overlaps}
\end{equation}
where ${\cal A}$ is the cross-sectional area of the wire.
As in Eqs. (\ref{asymr}) and (\ref{asymr2}), the $L$
here refers to scattering states which start on the left,
$k_z>0$, and $R$ refers to those which start on the right,
$k_z<0$.  The distinction between left and right moving
is a probably unphysical here; however, it is useful to make contact
to the tunnel junction case.  Eq. (\ref{overlaps}) implies that
\begin{equation}
t^{R(L)}_{\bf k}  = \frac{W^{L(R)}e^{-ik\cdot a}}{\sqrt{{\cal A}}}
                  \frac 1{i|v_z|}.
\end{equation}
Finally, we define the scattering rate of state ${\bf k}$ 
from the impurity as
\begin{equation}
\Gamma ^A_{\bf k} = \frac {|W^A|^2}{{\cal A}}
                    \int \frac {dk_z}{2\pi} 
                    \pi\delta (\omega -\epsilon _{{\bf k}_\perp} 
                                       -\epsilon _{k_z})
                  = \frac 12\frac 1{|v_z|}
                    \frac {|W^A|^2}{{\cal A}},
\end{equation}
where the integral is done either over $k_z>0$ or $k_z<0$ for $A=L,\, R$,
respectively.  
Current conservation requires that $I_L=I_R$ so in computing our final result
for the current we can take any linear combination of $I_L$ and $I_R$
which is convenient:  
\begin{eqnarray}
I &=& \frac {|W^L|^2}{|W^L|^2+|W^R|^2}I_L
   +  \frac {|W^R|^2}{|W^L|^2+|W^R|^2}I_R \nonumber \\
  &=& \sum _{{\bf k}_{\perp}}
            \int \frac {d\omega}{2\pi} 
                (f_L(\omega ) - f_R(\omega )) \label{finalcur}
	     \\
   &-& \int {d\omega}
       \sum _{{\bf k}_{\perp}}
            \frac {2\Gamma _{\bf k}^L\Gamma _{\bf k}^R}
                  {\Gamma _{\bf k}^L+\Gamma _{\bf k}^R}
            A_d(\omega)
            (f_L(\omega) - f_R(\omega )) \nonumber ,
\end{eqnarray}
where $A_d(\omega ) = -\mbox{Im} g_r(0,0)/\pi$ is the impurity spectral function.

This is our final result for the number current.
The first term gives the Sharvin point contact conductance.
The second term is the correction to the current due to
the presence of the impurity.  The expression is the
same as for a tunnel junction,\cite{sdw,mewin2} except the sign is reversed.
The correction to the current once one includes the electron
spin is
\begin{equation}
\delta I = -  \int {d\omega}
       \sum _{{\bf k}_{\perp},s}
            \frac {2\Gamma _{{\bf k},s}^L\Gamma _{{\bf k},s}^R}
                  {\Gamma _{{\bf k},s}^L+\Gamma _{{\bf k},s}^R}
            A_{d,s}(\omega )
            (f_L(\omega) - f_R(\omega )),
\end{equation}
where the only change from Eq. (\ref{finalcur}) is that there is a sum
over the spin, $s$. If we assume a constant density of states and no spin
dependence of the matrix elements,
we obtain the expression Eq. (\ref{curr}), except for the difference
in sign between the tunnel junction and point contact case.
Note that in Eq. (\ref{curr}) the $\Gamma _L$ and $\Gamma _R$
are defined with the density of states divided out.

\vspace*{0.2cm}
%

%
%
\newpage
%
\end{document}